\documentclass[useAMS,usenatbib]{mn2e_personal}

\usepackage{graphicx}

\title[Implicit integrations in SPH modelling]{Implicit integrations for SPH in Semi-Lagrangian approach: application to the accretion disc modelling in a microquasar}
\author[G. Lanzafame]{G. Lanzafame\thanks{E-mail:
glanzafame@oact.inaf.it}\\
INAF - Osservatorio Astrofisico di Catania, Via S. Sofia
              78 - 95123 Catania, Italy\\}
\begin{document}

\date{Accepted -------. Received -------; in original form -------}

\pagerange{\pageref{firstpage}--\pageref{lastpage}} \pubyear{2009}

\maketitle

\label{firstpage}

\begin{abstract}
Current explicit integration techniques in fluid dynamics are deeply limited by the Courant-Friedrichs-Lewy condition of the time step progression, based on the adopted spatial resolution coupled with the maximum value between the kinetic velocity or the signal transmission speed in the computational domain. Eulerian implicit integration techniques, even though more time consuming, can allow to perform stable computational fluid dynamics paying the price of a relatively larger inaccuracy in the calculations, without suffering such a strict temporal limitation. In this paper, we present a simple and effective scheme to perform Free Lagrangian Smooth Particle Hydrodynamics (SPH) implicit integrations in Semi-Lagrangian approach without any Jacobian matrix inversion operations for viscous Navier-Stokes flows. Applications to SPH accretion disc simulation around a massive black hole (MBH) in a binary stellar system are shown, together with the comparison to the same results obtained according to the traditional explicit integration techniques. Some 1D and 2D critical tests are also discussed to check the validity of the technique.
\end{abstract}


\section{Introduction}

  A time step restriction is always necessary for time dependent calculations in computational fluid dynamics. Currently, such restrictions are needed for mathematical stability reasons in explicit calculations of partial differential equations (PDE), while they are necessary for accuracy considerations in implicit calculations. The integrated physical local property at time step level $n+1$: $A^{n+1}$ is a function of its previous values at time steps $n$, $n-1$ etc., as well as of spatial derivatives of its spatial flux densities: $dF(A)/dr$, relative to the previous time steps for explicit calculations. Instead $A^{n+1}$ is a function of these quantities also for the same $n+1$ time step level for implicit calculations.

  For explicit calculations of the computational flow, the Courant-Friedrichs-Lewy condition \citep{c1,c2} is imposed on those hyperbolic terms representing advection in PDE (spatial derivatives of pressure or velocity), where the given Courant number $C = v_{c} \Delta t_{CFL}/\Delta r \leq 1$ is generally of the order of $0.2 - 0.5$, where $\Delta r$ is the spatial resolution, $v_{c}$ is the maximum value between the local kinematic and the signal transmission velocities within the entire computational domain, and $\Delta t_{CFL}$ is the Courant-Friedrichs-Lewy time step to be computed.

  Explicit integration techniques are widely adopted to solve equations of the fluid dynamics both in the Eulerian formalism, where time and space derivatives refer both to local derivatives of the physical properties ($\partial/\partial t$ and $\partial/\partial r$), according to the adopted spatial resolution length, and in the Lagrangian formalism, where the material or the convective derivative $d/dt = \partial/\partial t + \bmath{v} \cdot \nabla$ characterizes the flow description \citep{c3,c4,c5}. Although the scientific literature is quite rich as for the implicit numerical integrations of PDE, implicit backward difference methods are also shown for ordinary differential equations (ODE) working with Lagrangian derivatives \citep{c70,c71}. In spite of such a dichotomy \citet{c6} produced an implicit Lagrangian method, based on a triangular mesh for calculations of non stationary astrophysical processes, whose results successfully compare with other Lagrangian explicit calculations, although some details in the flow do not compare with those obtained via Eulerian formalism. In Semi-Lagrangian techniques, the basic idea is to discretize the Lagrangian derivative of the solution in time instead of the Eulerian derivative. Despite the full Lagrangian nature of the technique, a spatial grid is necessarily considered. Values of physical quantities of the system of differential equations are calculated through spatial interpolations at the grid points. Explicit-implicit integration techniques involve these spatial grid points and particle positions at previous time levels. Semi-Lagrangian schemes \citep{c21,c22} have been built up based on mixed explicit-implicit schemes, increasing the time step up to a factor of $6$, paying a little additional cost in the time computing, with a limited degradation in the accuracy of solutions. However, the evaluation of the maximum stable time step still remains debated because a time step lengthening of a factor of $6$ appears much shorter than seems necessary from considerations of accuracy \citep{c23}. These works of the 70's - 80's were developed in metereology for numerical weather forecasts, where the use of a longer time step is essential for efficiency. 2D and 3D applications of integrations in Semi-Lagrangian approach have been discussed in \citet{c20} following a hierarchy of cases.

  The extension of Semi-Lagrangian method to the solution of viscous Navier-Stokes equations was done by \citet{b13}, showing the nonlinear stability of the integration method, even as the viscosity decreases to zero, also showing an estimate of errors, later improved by \citet{b14,b15}.

  Currently, SPH hydrodynamics works adopting an explicit integration technique, being SPH a "Free Lagrangian particle scheme" \citep{c54}. Recently some authors \citep{c10,c9,c8,c7} developed implicit integration schemes working in SPH to solve some selected problems dealing with the radiative transfer or with the heat transfer in the flow, adopting either the Crank-Nicolson, or the Runge-Kutta-Fehlberg numerical integrations, or the conjugate gradient method. The adoption of such methods involves the handling of some time-expensive and memory consuming Jacobian matrices. Alternatively \citep{c55}, some iterative variational techniques are also used to find zeros for the analytical equation coming out from the energy equation written in finite terms and including the diffusive terms, within an assigned tolerance error, setting the "left" and "right" boundary values for the thermal energy per unit mass $\epsilon$, finding the median value using the Van Wijingarden-Dekker-Brent bisection method \citep{c56}. However, the introduction of such mixed procedures involves deviations from the original pure particle hydrodynamics.

  In this paper we present a simple implicit mathematical technique able to perform Semi-Lagrangian explicit-implicit integrations both of the Euler and of the Navier-Stokes equations of the fluid dynamics, respecting the SPH, or the SPH-derived schemes, without any Jacobian matrix handling.

  The implicit integration procedure in our scheme is naturally based on an iterative integration scheme. However, we pay attention to the solution of the entire system of equations for all SPH particles at the same time instant without using interpolation techniques like other Semi-Lagrangian schemes need \citep{c21,c22}, limiting the procedure to the implicit solution of a single advection equation even in 3D \citep{c20}. Instead, we use SPH interpolations to perform implicit iterative integrations on the entire system of equations for all SPH particles within the computational domain.

  An implicit integration scheme promises to be more efficient in some situations, for example when large bulk velocities are present, or when a system is close to hydrostatic equilibrium, or for very low Mach number flows, for string diffusion, or for stiff source terms, etc.. In such cases the Courant-Friedrichs-Lewy time step constraint that needs to be obeyed in explicit time integration schemes can be very restrictive, and may yield an explicit time step much shorter than in principle needed to follow the real dynamics of the system.

  The next sections of this work show how a Semi-Lagrangian explicit-implicit technique can be applied to SPH. To do this, we briefly recall how SPH works in \S 2. Then, we show how implicit integrations schemes work, how the Lagrangian Euler or Navier-Stokes equations have to be rewritten, how they are translated in the finite-difference schemes and how a suitable implicit time step advancement well beyond the Courant-Friedrichs-Lewy limit is chosen (\S 3). Successful applications are here also reported for the case of 1D and 2D blast waves SPH modelling, as well as for the 2D gas shockless radial viscous transport (\S 4). Finally, astrophysical applications are here also presented (\S5) on the study of a 3D an accretion disc around a MBH in a microquasar, both adopting the implicit and the explicit integration procedures. Some advantages and some disadvantages of the implicit techniques, compared with the explicit schemes are also discussed. Both explicit and implicit accuracies, applied to SPH, are also discussed.

  In a physically viscous accretion process onto a MBH in a microquasar, the viscous conversion of mechanical energy into heat decreases the Mach number to values so low that the CFL explicit time step could be too short for practical purposes.

\section{Fluid dynamics equations and their SPH formulation}

  The relevant equations to our model or viscous gas hydrodynamics are:

\begin{equation}
\frac{d\rho}{dt} + \rho \nabla \cdot \bmath{v} = 0 \hfill \mbox{continuity equation}
\end{equation}

\begin{eqnarray}
\frac{d \bmath{v}}{dt} & = & - \frac{\nabla p}{\rho} + \bmath{f} + \frac{1}{\rho} \nabla \cdot \bmath{\tau} \nonumber \\
& &  \ \ \ \ \ \hfill \mbox{Navier-Stokes momentum equation}
\nonumber
\end{eqnarray}

\begin{eqnarray}
\frac{d}{dt} \left( \epsilon + \frac{1}{2} v^{2}\right) = - \frac{1}{\rho} \nabla \cdot \left( p \bmath{v} - \bmath{v} \cdot \bmath{\tau} \right) + \bmath{f} \cdot \bmath{v} \nonumber
\end{eqnarray}

\begin{equation}
\hfill \mbox{energy equation}
\end{equation}

\begin{equation}
p = f(\gamma, \rho, \epsilon, \bmath{r}, \bmath{v}) \hfill \mbox{equation of state}
\end{equation}

\begin{equation}
\frac{d \bmath{r}}{dt} = \bmath{v} \hfill \mbox{kinematic equation.}
\end{equation}

  $d/dt$ stands for the Lagrangian derivative, $\rho$ is the gas density, $\epsilon$ is the thermal energy per unit mass, $\bmath{f}$ is an external force field, $p$ is the ideal gas pressure, here generally expressed as a function of local properties, determined by its equation of state (EoS). The adiabatic index $\gamma$ has the meaning of a numerical parameter whose value lies in the range between $1$ and $5/3$, in principle. $\bmath{\tau}$ is the viscous stress tensor, whose presence modifies the Euler equations for a non viscous fluid dynamics in the viscous Navier-Stokes equations. Notice that the inclusion of the field terms (inertial or non inertial) does not affect the mathematical scheme regarding the theme of this paper. Sometimes these terms are absent and sometimes they are present, according the simulation we are considering. Here they are written only for universality reasons.

  The SPH method is a Free Lagrangian scheme \citep{c54} that discretizes the fluid into moving interacting and interpolating domains called "particles" \citep{c11,c12,c13}. All particles move according to pressure and body forces. The method makes use of a Kernel $W$ useful to smoothing interpolate a physical quantity $A(\bmath{r})$ related to a gas particle at position $\bmath{r}$ according to:

\begin{equation}
\overline{A}(\bmath{r}) = \int_{D} A(\bmath{r}') W(\bmath{r}, \bmath{r}', h) d \bmath{r}'.
\end{equation}

$W(\bmath{r}, \bmath{r}', h)$, the interpolation Kernel, is a continuous function - or two connecting continuously differentiable functions even at the connecting point - defined in the spatial range $2h$, whose limit for $h \rightarrow 0$ is the Dirac delta distribution function. All physical quantities are described as extensive properties smoothly distributed in space and computed by interpolation at $\bmath{r}$. In SPH terms we write:

\begin{equation}
\overline{A}_{i} = \sum_{j=1}^{N} \frac{A_{j}}{n_{j}} W(\bmath{r}_{i}, \bmath{r}_{j}, h) = \sum_{j=1}^{N} \frac{A_{j}}{n_{j}} W_{ij}
\end{equation}

where the sum is extended to all particles included within the domain $D$, $n_{j} = \rho_{j}/m_{j}$ is the number density relative to the jth particle. $W(\bmath{r}_{i}, \bmath{r}_{j}, h)$ is the adopted interpolation Kernel whose value is determined by the relative distance between particles $i$ and $j$. $\int W(\bmath{r}_{i}, \bmath{r}_{j}, h) d^{3} \bmath{r}' = 1$, that is: $\sum_{j} W(\bmath{r}_{i}, \bmath{r}_{j}, h)/n_{j} = 1$.

  In SPH conversion of mathematical equations there are two principles embedded. Each SPH particle is an extended, spherically symmetric domain where any physical quantity $f$ has a density profile $f W(\bmath{r}_{i}, \bmath{r}_{j}, h) \equiv f W(|\bmath{r}_{i} - \bmath{r}_{j}|, h) = f W(|\bmath{r}_{ij}|,h)$. Besides, the fluid quantity $f$ at the position of each SPH particle could be interpreted by filtering the particle data for $f(\bmath{r})$ with a single windowing function whose width is $h$. So doing, fluid data are considered isotropically smoothed all around each particle along a length scale $h$. Therefore, according to these two concepts, the SPH value of the physical quantity $f$ is both the overlapping of extended profiles of all particles and the overlapping of the closest smooth density profiles of $f$. This means that the compactness of the Kernel shape gives the principal contribution to the interpolation summation to each particle by itself and by its closest neighbours. In both approaches the mass is globally conserved in so far as the total particle number is constant.

  In SPH formalism, equations (2) and (3) take the form:

\begin{eqnarray}
\frac{d \bmath{v}_{i}}{dt} & = & - \sum_{j=1}^{N} m_{j} 
\left( \frac{p_{i}^{\ast}}{\rho_{i}^{2}} + \frac{p_{j}^{\ast}}{\rho_{j}^{2}} \right) \nabla_{i} W_{ij} + \bmath{g}_{i} + \nonumber \\
& & \sum_{j=1}^{N} m_{j} \left( \frac{\eta_{vi} \mbox{\boldmath  $\sigma$}_{i}}{\rho_{i}^{2}} + \frac{\eta_{vj} \mbox{\boldmath $\sigma$}_{j}}{\rho_{j}^{2}} \right) \cdot \nabla_{i} W_{ij} \\
\frac{d}{dt} E_{i} & = & - \sum_{j=1}^{N} m_{j} \left( \frac{p_{i}^{\ast} \bmath{v}_{i}}{\rho_{i}^{2}} + \frac{p_{j}^{\ast} \bmath{v}_{j}}{\rho_{j}^{2}}\right) \cdot \nabla_{i} W_{ij} + \bmath{g}_{i} \cdot \bmath{v}_{i} + \nonumber \\
& & \sum_{j=1}^{N} m_{j} \left( \eta_{vi} \frac{\mbox{\boldmath $\sigma$}_{i} \cdot \bmath{v}_{i}}{\rho_{i}^{2}} + \eta_{vj} \frac{\mbox{\boldmath $\sigma$}_{j} \cdot \bmath{v}_{j}}{\rho_{j}^{2}} \right) \cdot \nabla_{i} W_{ij}
\end{eqnarray}

where $m_{j}$ is the mass of jth particle and $p_{i}^{\ast} = p_{i} +$ {\it dissipation pressure term}. $E_{i} = (\epsilon_{i} + \frac{1}{2} v_{i}^{2})$. The viscous stress tensor $\tau_{\alpha \beta}$ includes the positive first and second viscosity coefficients $\eta_{v}$ and $\zeta_{v}$ which are velocity independent and describe shear and tangential viscosity stresses ($\eta_{v}$), and compressibility stresses ($\zeta_{v}$):

\begin{equation}
\tau_{\alpha \beta} = \eta_{v} \sigma_{\alpha \beta} + \zeta_{v} \nabla 
\cdot \bmath{v}
\end{equation}

where the shear

\begin{equation}
\sigma_{\alpha \beta} = \frac{\partial v_{\alpha}}{\partial x_{\beta}} 
+ \frac{\partial v_{\beta}}{\partial x_{\alpha}} - \frac{2}{3} 
\delta_{\alpha \beta} \nabla \cdot \bmath{v}
\end{equation}

  In these equations $\alpha$ and $\beta$ are spatial indexes while tensors are written in bold characters. For the sake of simplicity we assume $\zeta_{v} = 0$, however our code allows us also different choices. Defining

\begin{equation}
V_{i \alpha \beta} = \sum_{j=1}^{N} \frac{m_{j} \bmath{v}_{ji 
\alpha}}{\rho_{j}} \frac{\partial W_{ij}}{\partial x_{\beta}}
\end{equation}

as the SPH formulation of $\partial v_{\alpha}/\partial x_{\beta}$, the SPH equivalent of the shear is:

\begin{equation}
\sigma_{i \alpha \beta} = V_{i \alpha \beta} + V_{i \beta \alpha} - \frac{2}{3} \delta_{\alpha \beta} V_{i \gamma \gamma}
\end{equation}

A full justification of this SPH formalism can be found in \citet{c14,c15}.

In this scheme the continuity equation takes the form:

\begin{equation}
\frac{d\rho_{i}}{dt} = \sum_{j=1}^{N} m_{j} \bmath{v}_{ij} \cdot 
\nabla_{i} W_{ij}
\end{equation}

or, as we adopt, it can be written as:

\begin{equation}
\rho_{i} = \sum_{j=1}^{N} m_{j} W_{ij}
\end{equation}

which identifies the natural space interpolation of particle densities according to equation (7).

  In such a conversion, the physical mass density of the $i$th SPH particle is either physically represented as $\rho_{i} = m_{i} n_{i}$, or numerically represented as $\rho_{i} = \sum_{j} \rho_{j}/n_{j} W_{ij} = \sum_{j} m_{j} W_{ij}$. In the same fashion, the particle numerical density is either $n_{i} = \rho_{i}/m_{i} $, or $n_{i} = \sum_{j} W_{ij}$.

  A necessary convergence is needed, because the two expressions for $\rho_{i}$ or for $n_{i}$ coincide only in the case of equal mass SPH particles, when $m_{i} = m_{j}$.

  In addition, looking at the SPH particle masses, either $m_{i} = \rho_{i}/n_{i}$, or $m_{i} = \sum_{j} m_{j}/n_{j} W_{ij} = \sum_{j} \rho_{j}/n_{j}^{2} W_{ij}$, which equal with each other only when $n_{i} = n_{j}$.

  In some circumstances SPH particles could have different physical masses. In this case, the concept of SPH particle mass could be without any physical meaning whenever local densities are different: $\rho_{i} \neq \rho_{j}$.

  To resolve this incongruity, we modify the SPH formulation (15) according to one of these choices:

either

\begin{equation}
\left\{ \begin{array}{lll}
m_{i} = \frac{\rho_{i}}{n_{i}} = \frac{1}{n_{i}} \sum_{j} \frac{\rho_{j}}{n_{j}} W_{ij} \\
n_{i} = \frac{\rho_{i}}{m_{i}} = \frac{1}{m_{i}} \sum_{j} m_{j} W_{ij} \\
\rho_{i} = m_{i} n_{i} = \sum_{j} m_{j} W_{ij},
\end{array} \right.
\end{equation}

as we did in this paper, or

\begin{equation}
\left\{ \begin{array}{lll}
m_{i} = \frac{\rho_{i}}{n_{i}} = \sum_{j} \frac{m_{j}}{n_{j}} W_{ij} = \sum_{j} \frac{\rho_{j}}{n_{j}^{2}}W_{ij} \\
n_{i} = \frac{\rho_{i}}{m_{i}} = \frac{\rho_{i}}{m_{i}} \sum_{j} \frac{m_{j}}{\rho_{j}} W_{ij} \\
\rho_{i} = m_{i} n_{i} = n_{i} \sum_{j} \frac{m_{j}}{n_{j}} W_{ij},
\end{array} \right.
\end{equation}

or

\begin{equation}
\left\{ \begin{array}{lll}
m_{i} = \frac{\rho_{i}}{n_{i}} = \frac{\rho_{i}}{n_{i}} \sum_{j} \frac{1}{n_{j}} W_{ij} \\
n_{i} = \frac{\rho_{i}}{m_{i}} = \sum_{j} W_{ij} \\
\rho_{i} = m_{i} n_{i} = m_{i} \sum_{j} W_{ij}.
\end{array} \right.
\end{equation}

  The first of these $\rho - n - m$ correlations (eqs. 16) is written considering the SPH interpolation integral (eqs. 6, 7) applied only to the mass density $\rho$, not to $m$ or to $n$. The second correlation (eqs. 17) is obtained considering eqs. (6, 7) applied only to the mass $m$, not to $\rho$ or to $n$; instead the third form (eqs. 18) comes out applying eqs. (6, 7) only to the numerical density $n$, not to $\rho$ or to $m$.

  Being $\rho$, $n$ and $m$ physically correlated as $\rho = m n$, the SPH interpolation integral (eqs. 6, 7) cannot be used at the same time for all these three quantities, otherwise differences among smoothed to unsmoothed quantities could be relevant.

  Instead, according to one of eqs. 16, 17 or 18 choices, the full convergence between the SPH transformations and the full physical meaning of mass and densities are respected.

  In its original SPH formulation, standing the ideal equation of state (EoS) in the form:

\begin{equation}
p = (\gamma - 1) \rho \epsilon \hfill \mbox{perfect gas equation,}
\end{equation}

dissipation in $p^{\ast}$ is given by an artificial viscosity term. This dissipation, together with an appropriate numerical thermal diffusion contribution $\propto (U_{j} - U_{j}) c_{sij}/\rho_{ij}$, where $U_{i} = \rho_{i} \epsilon_{i}$ \citet{c11,c12,c13}, included in $d \epsilon/dt$, reduce shock fluctuations. The artificial viscosity term is given by:

\begin{equation}
\eta_{ij} = \alpha_{SPH} \mu_{ij} + \beta_{SPH} \mu_{ij}^{2},
\end{equation}

where

\begin{equation}
\mu_{ij} = \left\{ \begin{array}{ll}
\frac{2 h \bmath{v}_{ij} \cdot \bmath{r}_{ij}}{(c_{si} + c_{sj}) (r_{ij}^{2} + \xi^{2})} & \textrm{if $\bmath{v}_{ij} \cdot \bmath{r}_{ij} < 0$}\\
\\
0 & \textrm{otherwise}
\end{array} \right.
\end{equation}

being  $c_{si}$ the sound speed of the ith particle, ${\bmath r}_{ij} = {\bmath r}_{i} - {\bmath r}_{j}$, ${\bmath v}_{ij} = {\bmath v}_{i} - {\bmath v}_{j}$, $\xi^{2} \ll h^{2}$, $\alpha_{SPH} \approx 1$, $\beta_{SPH} \approx 2$, $c_{sij} = 0.5 (c_{si} + c_{sj})$ and $\rho_{ij} = 0.5 (\rho_{i} + \rho_{j})$. These $\alpha_{SPH}$ and $\beta_{SPH}$ parameters of the order of the unity are usually adopted to damp oscillations past high Mach number shock fronts developed by non-linear instabilities \citep{c18}. These $\alpha_{SPH}$ and $\beta_{SPH}$ values were also adopted by \citet{c17}. Smaller $\alpha_{SPH}$ and $\beta_{SPH}$ values, as adopted by \citet{c19}, for developing more turbulence. In the physically inviscid SPH gas dynamics, angular momentum transport is mainly due to the artificial viscosity included in the pressure terms as:

\begin{equation}
\frac{p_{i}^{\ast}}{\rho_{i}^{2}} + \frac{p_{j}^{\ast}}{\rho_{j}^{2}} = \left( \frac{p_{i}}{\rho_{i}^{2}} + \frac{p_{j}}{\rho_{j}^{2}} \right) (1 + \eta_{ij})
\end{equation}

where $p$ is the intrinsic gas pressure.

  However, looking at a physical origin of the numerical dissipation term, included in the EoS \citep{c16,c24,c25}, gas pressure can be expressed as:

\begin{equation}
p^{\ast} = \frac{\rho}{\gamma} c_{s}^{2} \left(1 - C \frac{n^{-1/3} \nabla \cdot \bmath{v}}{3 c_{s}} \right)^{2},
\end{equation}

where

\begin{equation}
C = \frac{1}{\pi}  \textrm{arccot} \left( R \frac{v_{R}}{c_{s}} \right),
\end{equation}

where $R \gg 1$. $R$ is a large number describing how much the flow description corresponds to that of an ideal gas: $R \approx \lambda/d$, being $\lambda \propto \rho^{-1/3}$ the molecular mean free path, and being $d$ the mean linear dimension of gas molecules. $v_{R}$ is the impact relative velocity component. The physical dissipation, expressed by the two further terms in eq. (23) (the linear and the quadratic terms in $\nabla \cdot \bmath{v}$) of the reformulated EoS, better treats both shocks and shear flows, even in a Lagrangian description. Their inclusion substitutes artificial viscosity terms and does not represent a physical turbulent viscous contribution, but the physical dissipation coming out because eq. (19) of the EoS should strictly be applied only to quasi-static processes \citep{c16,c24,c25}.

\section{Concepts on explicit and implicit integration schemes: application to SPH}

\subsection{Formulations on explicit and implicit generalized three-level integration schemes}

  Given a physical property $A$, the mathematical conversion of its time derivative $\partial A/\partial t$ and of its first and second space derivatives: $\partial A/\partial r$ and $\partial^{2} A/\partial r^{2}$ is: $\partial A/\partial t \rightarrow (A_{k}^{n+1} - A_{k}^{n})/\Delta t$, $\partial A/\partial r \rightarrow 0.5 (A_{k+1}^{n} - A_{k-1}^{n})/\Delta r_{k}$ and $\partial^{2} A/\partial r^{2} \rightarrow 0.5 (A_{k+1}^{n} - 2 A_{k}^{n} + A_{k-1}^{n})/\Delta r_{k}$, respectively for explicit techniques, while it is $\partial A/\partial t \rightarrow (A_{k}^{n+1} - A_{k}^{n})/\Delta t$, $\partial A/\partial r \rightarrow 0.5 (A_{k+1}^{n+1} - A_{k-1}^{n+1})/\Delta r_{k}$ and $\partial^{2} A/\partial r^{2} \rightarrow 0.5 (A_{k+1}^{n+1} - 2 A_{k}^{n+1} + A_{k-1}^{n+1})/\Delta r_{k}$, respectively for implicit techniques \citep{c3,c4,c5}. In such expressions, $n$ represents the temporal level, while $k$ represents the spatial grid index, according to the versus of the reference frame. $\Delta t = t^{n+1} - t^{n}$ and $\Delta r_{k} = r_{k+1} - r_{k} = r_{k} - r_{k-1} = 0.5 (r_{k+1} - r_{k-1})$. Of course other higher order schemes exist, where more time and space levels are considered \citep{c3,c4,c5}, which we do not consider for the sake of simplicity.

  As a useful generalization \citep{c3}, the hypothetical equation:

\begin{equation}
\frac{\partial A}{\partial t} + a \frac{\partial A}{\partial r} + b \frac{\partial^{2} A}{\partial r^{2}} + c = 0
\end{equation}

can be converted as:

\begin{equation}
\left\{ \begin{array}{llll}
(1 + \lambda) \frac{A_{k}^{n+1} - A_{k}^{n}}{\Delta t} - \lambda \frac{A_{k}^{n} - A_{k}^{n-1}}{\Delta t} + \\
a \left[ (1 - \kappa) \frac{A_{k+1}^{n} - A_{k-1}^{n}}{2 \Delta r_{k}} + \kappa \frac{A_{k+1}^{n+1} - A_{k-1}^{n+1}}{2 \Delta r_{k}} \right] + \\
b \left[ (1 - \kappa) \frac{A_{k+1}^{n} - 2 A_{k}^{n} + A_{k-1}^{n}}{2 \Delta r_{k}} + \kappa \frac{A_{k+1}^{n+1} - 2 A_{k}^{n+1} + A_{k-1}^{n+1}}{2 \Delta r_{k}} \right] + \\
c = 0.
\end{array} \right.
\end{equation}

  Whenever the pair $\lambda = 0$ and $\kappa = 0$, such a general expression yields the typical explicit two points forward centred integration scheme (2FCS) for PDE. Instead, for $\lambda = 0$ and $\kappa = 1$ we have a simple full implicit centred three points integration scheme (3FICS). Furthermore, for $\lambda = 0.5$ and $\kappa = 1$ we have a linearized full implicit three points technique (3LFI), while for $\lambda = 0$ and $\kappa = 0.5$ we obtain the well known Crank-Nicholson implicit scheme \citep{c3,c4,c5}.

  The truncation error \citep{c3} is a function $\propto \Delta r f(\partial^{2} A/\partial r^{2}, \partial^{3} A/\partial r^{3})$ for upwind 2FCS, as well as for 3FICS schemes, while it is $\propto \Delta r^{2} f(\partial^{3} A/\partial r^{3}, \partial^{4} A/\partial r^{4})$ for the other last two implicit techniques.

  Although the expressions (25) and (26) refer to the numerical value of a non-smoothed or intrinsic physical property $A$, we assume that it also holds for its smoothed SPH evaluation. Hence we can rewrite the same expressions using indifferently $\overline{A}$ instead of $A$ in eqs. (6) and (7). Since we integrate both explicitly and implicitly only smoothed physical quantities, throughout the rest of the paper we will use mathematical symbols without any superscript line for a simpler reading.

\subsection{SPH in a Semi-Lagrangian explicit-implicit generalized three-level integration scheme}

  Hydrodynamics in the nonlinear Free Lagrangian SPH approach is currently performed in predictor-corrector explicit schemes, starting from some initial values at time $t = 0$. In the "Leapfrog" scheme, the equations for space and velocity advancement can be written as:

\begin{eqnarray}
r_{k}^{n+1} & = & r_{k}^{n} + v_{k}^{n+1/2} \Delta t \\
v_{r,k}^{n+1/2} & = & v_{r,k}^{n-1/2} + a_{r,k}^{n} \Delta t
\end{eqnarray}

that can be manipulated into a form which writes particle velocity at integer steps as

\begin{eqnarray}
r_{k}^{n+1} & = & r_{k}^{n} + v_{r,k}^{n} \Delta t + \frac{1}{2} a_{r,k}^{n} \Delta t^{2} \\
v_{r,k}^{n+1} & = & v_{r,k}^{n} + \frac{1}{2} (a_{r,k}^{n} + a_{r,k}^{n+1}) \Delta t.
\end{eqnarray}

  In this second expression, since particle acceleration $a$ depends on $v$, it is required an implicit integration for the second equation. In the case of a "Leapfrog" scheme, an "evaluator" phase in the computational scheme needs to be interposed between the two integration procedures, where time derivatives of the various physical quantities are computed. For this reason, this scheme is a so called PEC method, where a Predictor-Evaluator-Corrector procedure is followed by the updating of all integrated values.

  Iterative Runge-Kutta methods are also used, both explicit as well as implicit. In such schemes, the integrated value of the physical property $A_{k}^{n+1} = A_{k}^{n} + S \Delta t$, where $S$ is an estimated weighted average of slopes from the beginning, through some midpoints, toward the end of the time interval. Despite more general than explicit methods, implicit Runge-Kutta schemes are more complicated and dependent on the specific problem. Those Runge-Kutta methods that are diagonally implicit, show a strong stability allowing a significant increase in the time step limit, compared with the explicit methods of the same order \citep{c27,c26}. Here, we do not discuss in detail this specific complex argument, where often some Jacobian matrices need to be inverted. However, even the simple backward Euler ($\lambda = 0$, $\kappa = 1$), or the Crank-Nicholson ($\lambda = 0$, $\kappa = 0.5$) methods belong to this category.

  The explicit multistep Adams-Bashforth-Moulton PECE explicit integration scheme can also be adopted in SPH, where either the Adams-Bashforth

\begin{figure}
\resizebox{\hsize}{!}{\includegraphics[clip=true]{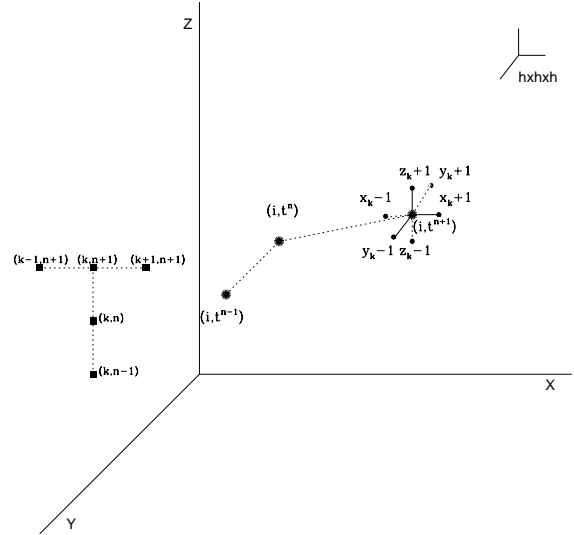}}
\caption{Schematic plot showing the position of the ith SPH particle at time $t^{n-1}$, $t^{n}$ and $t^{n+1}$. At time $t^{n+1}$ the X,Y,Z spatial grid is shown, where physical properties are interpolated, used to perform the implicit 3FICS scheme, together with the information relative to the same particle at previous times. A stencil diagram referring to the 1D implicit integration scheme is also reported on the left of the same picture showing the spatial $k$ and temporal $n$ indexes. In 3D, a distorted stencil scheme is used in like manner as in 1D.}
\end{figure}

\begin{equation}
\left\{ \begin{array}{llll}
A_{k}^{n+1} = A_{k}^{n} + \Delta t \frac{\partial A_{k}^{n}}{\partial t} \\
A_{k}^{n+2} = A_{k}^{n+1} + \frac{\Delta t}{2} \Bigg( 3 \frac{\partial A_{k}^{n+1}}{\partial t} - \frac{\partial A_{k}^{n}}{\partial t} \Bigg) \\
A_{k}^{n+3} = A_{k}^{n+2} + \frac{\Delta t}{12} \Bigg( 23 \frac{\partial A_{k}^{n+2}}{\partial t} - 16 \frac{\partial A_{k}^{n+1}}{\partial t} + 5 \frac{\partial A_{k}^{n}}{\partial t} \Bigg) \\
A_{k}^{n+4} = A_{k}^{n+3} + \frac{\Delta t}{24} \Bigg( 55 \frac{\partial A_{k}^{n+3}}{\partial t} - 59 \frac{\partial A_{k}^{n+2}}{\partial t} + \\ 37 \frac{\partial A_{k}^{n+1}}{\partial t} - 9 \frac{\partial A_{k}^{n}}{\partial t} \Bigg) \\
A_{k}^{n+5} = A_{k}^{n+4} + \frac{\Delta t}{720} \Bigg( 1901 \frac{\partial A_{k}^{n+4}}{\partial t} - 2774 \frac{\partial A_{k}^{n+3}}{\partial t} + \\ 2616 \frac{\partial A_{k}^{n+2}}{\partial t} - 1274 \frac{\partial A_{k}^{n+1}}{\partial t} + 251 \frac{\partial A_{k}^{n}}{\partial t} \Bigg),
\end{array} \right.
\end{equation}

or the Adams-Moulton

\begin{equation}
\left\{ \begin{array}{llll}
A_{k}^{n} = A_{k}^{n-1} + \Delta t \frac{\partial A_{k}^{n}}{\partial t} \\
A_{k}^{n+1} = A_{k}^{n} + \frac{\Delta t}{2} \Bigg( \frac{\partial A_{k}^{n+1}}{\partial t} + \frac{\partial A_{k}^{n}}{\partial t} \Bigg) \\
A_{k}^{n+2} = A_{k}^{n+1} + \frac{\Delta t}{12} \Bigg( 5 \frac{\partial A_{k}^{n+2}}{\partial t} + 8 \frac{\partial A_{k}^{n+1}}{\partial t} - \frac{\partial A_{k}^{n}}{\partial t} \Bigg) \\
A_{k}^{n+3} = A_{k}^{n+2} + \frac{\Delta t}{24} \Bigg( 9 \frac{\partial A_{k}^{n+3}}{\partial t} + 19 \frac{\partial A_{k}^{n+2}}{\partial t} - \\ 5 \frac{\partial A_{k}^{n+1}}{\partial t} + \frac{\partial A_{k}^{n}}{\partial t} \Bigg) \\
A_{k}^{n+4} = A_{k}^{n+3} + \frac{\Delta t}{720} \Bigg( 251 \frac{\partial A_{k}^{n+4}}{\partial t} + \\ 646 \frac{\partial A_{k}^{n+3}}{\partial t} - 264 \frac{\partial A_{k}^{n+2}}{\partial t} + 106 \frac{\partial A_{k}^{n+1}}{\partial t} - 19 \frac{\partial A_{k}^{n}}{\partial t} \Bigg).
\end{array} \right.
\end{equation}

are used. To apply either the Adams-Bashforth or the Adams-Moulton techniques, up to the wished precision, previous derivatives for the same flow elements need to be conserved. Besides, a further evaluator procedure is considered at the end of the predictor-corrector integration scheme, to be a PECE (not a PEC) technique.

  What has been discussed is necessary to understand how it is possible to build up a Semi-Lagrangian explicit-implicit technique for SPH. In our Semi-Lagrangian explicit-implicit SPH integration approach, the first explicit integration scheme is necessarily the same that we currently use (Leapfrog, Adams-Bashforth-Moulton schemes, etc.), without strictly taking into account of the Courant-Friedrichs-Lewy limit for the time step. The adoption of an explicit-implicit backward and forward procedure has also been used by some others \citep{c30,c31,c28,c32,c29}, although some recent pure implicit schemes exist \citep{c33,c34,c35}, under some strict conditions. We use indifferently either a Leapfrog or an Adams-Bashforth-Moulton scheme as explicit procedure to calculate integrated values to the time level $n+1$ for all SPH particles. Then, we can use iteratively the expression (26) to repeat, only implicitly, integrations adopting $\lambda = 0.5$ and $\kappa = 1$, without any mathematical operation of Jacobian matrix inversion. The total number of implicit iterations is of course arbitrary. The first explicit integration step in the Semi-Lagrangian explicit-implicit technique works only once at the 1st iteration. From the 2nd iteration onward, the entire integration process is fully implicit. However, $2$ or $3$ implicit iterations at most are usually enough to get a convergence (even better in SPH) because the variation between the implicit solution and the previous solution in the iterative integration procedure quickly reduces toward very small relative differences $\Delta u/u$ (being $u = \rho, \epsilon, v_{X}, v_{Y}, v_{Z}$, avoiding any divergence in the solution \citep{b14,b15}. Notice that these relative differences refer to relative variations of the generic integrated physical quantity $u$ by the progression of iterative calculations of computed integrals, not to relative errors in the implicit integrations. Relative errors refer to relative discrepancies from the exact analytical solution, if it is known, or to general aspects referred to the entire flow through conservation laws. A number of implicit iterations larger than $3$ is of course possible, but practicable in so far as the implicit iteration technique is competitive against the current explicit integration methods. On the contrary, not only results would be less accurate, but also the entire procedure would be more time consuming. In this sense, the Semi-Lagrangian schemes of \citet{c21,c22}, whose method works for a time lengthening up to a factor of $6$, should be not convenient if implicit iterations are more than $4 - 5$. A quick convergence of numerical results in SPH is ensured by the fact that dissipation (even artificial) always works since the first explicit time step of the iterative cycle so that solutions normally relax without any oscillation or with very small damped oscillations. A larger number of iterations also produces a cumulative effect both on dissipation and on numerical errors, cycle by cycle, which is normally a not wished result. A control on the solution accuracy is normally possible in so far as only one SPH equation is implicitly integrated (as some authors do only for the energy equation). But, this is practically impossible when solving implicitly the entire system of equations, because how to control at the same time so many Semi-Lagrangian solutions as many are the scalar equations for a multitude of grid points is not treated in the mathematical literature, without any warranty for an improvement on other solutions. As an example, it could be possible that the implicit solution for the density for one SPH particle satisfies the arbitrary criterion that its relative variation is limited within the assigned tolerance as $\Delta \rho/\rho < 10^{-5}$ after $3$ implicit iterative cycles, but at the same time that the other relative variations for $\Delta \epsilon/\epsilon$, $\Delta v_{x}/v_{x}$, etc. still do not because they are greater than the established tolerance. The iterative cycle could either terminate, accepting the relative variations as they are at the moment, or it could continue. But each further cycle will destroy the consistency of the solution for the density which relaxes toward its numerical value within $3$ cycles of implicit iterations. From the 4th implicit cycle ahead the solution for $\rho$ could loose accuracy more and more. Even though normally a local instability affecting one solution (explicit or implicit) also affects other integrated solutions, situations where discrepancies occur in the relative variations cannot be excluded. These difficulties normally increase as longer is $\Delta t_{l}$ compared with $\Delta t_{CFL}$ because of both error and dissipation accumulation. Whenever $\Delta t_{l}/\Delta t_{CFL} < 6$, the degradation in the accuracy is normally negligible \citep{c21,c22}, but for larger time ratios it is not. This means that, especially for $10 < \Delta t/\Delta t_{CFL} < 20$, the number of iterations in the implicit procedure cannot be too low because the solution could be affected by the instabilities coming from the first explicit integration, but also it means at the same time that the number of iterations cannot be too high because of the long time lost and because of the inaccuracy corruption on the solution. A relaxation in the SPH solution normally happens. The density of energy $U = \frac{1}{2} \rho v^{2} + \rho \epsilon + \rho \bmath{f} \cdot \bmath{r}$ or the density of enthalpy $H = U + p$ include all physical quantities $\rho$, $v$, $\epsilon$, However, the propagation of relative discrepancies $\Delta u/u$ for both $\rho$, $\epsilon$, $v_{x}$, $v_{y}$ and $v_{z}$ cannot be managed cycle by cycle. What is here discussed clearly shows that the argument of solution consistency for a system of equations is quite complex for only one particle. However each iterative cycle of implicit integrations needs to involve a multitude of SPH particles at the same time. Relative variation for energy or enthalpy densities for the totality of particles as $\sum_{i}^{n} \Delta U_{i}/\sum_{i}^{n} U_{i}$ or as $\sum_{i}^{n} \Delta H_{i}/\sum_{i}^{n} H_{i} \leq $ an arbitrary tolerance (e.g. $10^{-5}$) is a valid general criterion for the evaluation of the consistency of the implicit integrated solutions at the same time level $n+1$ between two consecutive integration cycles. These global conditions do not ensure the consistency of local implicit solutions for the single particle even for explicit integrations within the CFL condition. This, in particular happens whenever free edges of the flow are considered, without any Dirichlet or Von Neuman boundary conditions. Indeed, two different SPH particles could have very different values for $\Delta U_{i}/U_{i}$ or for $\Delta H_{i}/H_{i}$, especially at the free edge of the computational domain. Hence, even if this global criterion is not satisfied at the 4th iterative cycle ($1$ explicit plus $3$ implicit), we decide to interrupt implicit iterations, proceeding to the next time step. Thus we privilege the criterion that the integration for each implicit iteration cannot exceed the 4th cycle (the third implicit cycle), otherwise the whole implicit scheme would be not competitive compared with the explicit integrations, because too time consuming. Since this paper is a study on the application of a Semi-Lagrangian integration technique to SPH, even the same comparison with SPH explicit integration solution should answer to the question whether the Semi-Lagrangian explicit-implicit integration consistency of solutions is acceptable or not.

  Our Semi-Lagrangian explicit-implicit integration technique on SPH costs about $10\%$ more time for each iterative implicit cycle (from the 2nd cycle onward) compared with a simple Leapfrog explicit scheme as used in SPH. This is due to the fact that the most expensive calculation is that relative only to the SPH interpolations of the two ends of each spatial stencil at time level $n + 1$ on the $i$th particle, fortunately restricted only to the same known local neighbours. This means that it is not necessary further computational time in searching for neighbours to the $i$th particle. At the same time, it is also necessary a limitation of the implicit dissipation accumulation. Truncation errors consist, typically, of successive higher even and odd ordered derivatives \citep{c3,c37,c4,c5} acting as a dissipation on the integrated solution. Therefore, each implicit iteration yields a solution including a dissipation, involving further stability, but also a larger inaccuracy, as it is normal for implicit integration calculations. It is normal that the implicit integrations are anyway less accurate than the explicit ones. However a much shorter duration of calculations should well compensate this disadvantage if consistency of solutions is satisfactory.

  Notice that in expression (26) while the $r$ subscript indicates the component of a vector determined by the orthogonal projection, the $k$ subscript is an index of position along a 1D spatial mesh at time ($n+1$), where SPH interpolation are needed. We adopt this scheme in 3D along three arbitrary distinct orthogonal axes (e.g. axes parallel to the $X$, $Y$, $Z$ directions), assuming that the real SPH particle stays at the centre of the spatial-temporal ($k$, $n+1$) 1D line parallel to the chosen arbitrary axis, time by time (Fig. 1), whose spatial grid index $k$ runs along it. The grid index $k$ identifies the ith particle along the 1D computational arbitrary spatial line (here $X, Y$, or $Z$) in 3LFI. Indexes $k+1$ and $k-1$ refer to the interpolation spatial points "ahead" and "behind" the same ith SPH particle along such an axis. The necessity of as many orthogonal axes as many are the dimensions of the flow is necessary to solve discontinuities in the flow whenever they are not simple linear fronts. So doing, we assume that both the Eulerian and the Lagrangian spatial derivatives are expressed in the same fashion in the finite differences scheme and that the correspondence between the temporal Lagrangian derivative $d/dt$ and its corresponding terms in the first row of eq. (26) still holds as for the Eulerian schemes. This can be done because such finite SPH discretizations refer to the moving Lagrangian particle, not to a spatial grid cell. On the left margin of Fig. 1, it is shown the space-time regular grid that is typically considered for 1D finite difference schemes \citep{c33,c34,c35} for a direct comparison to the distorted stencil scheme shown in 3D in the same picture. In reality, although the ($k$, $n$) space-time mesh term is technically correctly used, we do not use a real space-time mesh as for a pure Eulerian scheme in the 3D space, but simply some stencils connecting the ($k+1$, $n+1$), ($k$, $n+1$), ($k-1$, $n+1$) positions at time $n+1$ with the earlier ($k$, $n$), ($k$, $n-1$) positions of the ith SPH particle (Fig. 1).

  In the SPH approach, since smoothed solutions $\overline{A}$ are implicitly integrated instead of $A$ (\S 2, \S 3.1), as an indirect benefit, the convergence of solutions toward their values should be effective since the second implicit iteration. This normally does not suddenly happen for non smoothed physical solutions. Therefore, the Semi-Lagrangian explicit-implicit integration approach well combines with SPH in so far as the two pseudo-particles ahead ($k+1, n+1$) and behind ($k-1, n+1$) are located within $2h$ from the central real particle position at ($k, n+1$). This is the reason why we consider the spatial lengths from ($k-1, n+1$) to ($k, n+1$) and from ($k, n+1$) to ($k+1, n+1$) positions equal to $h$.

  The characteristic of Semi-Lagrangian explicit-implicit techniques is that the full Lagrangian description is conserved. The first step of the integration process is always explicit (as it is always necessary as for the full implicit techniques). Only physical values at the previous two time steps for each moving Lagrangian particle are needed. But, and this is the most important thing, only the physical information calculated at two adjacent fictional equally spaced pseudo particles along a stencil, where the real particle stays at its centre, in the 3D space are necessary. Hence a few space-time mesh points are involved for each iteration on each particle. This has the advantage that in the Semi-Lagrangian explicit-implicit integration process, heavy matrix inversion operations are avoided. Pure implicit integrations, using 3D grids are indeed memory and time consuming if a 3D simulation is considered, whose spatial resolution length is very short. The results are comparable, in principle, but the Semi-Lagrangian approach is technically faster and it requires a smaller computer memory.

  In the conversion of the physical Euler or Navier-Stokes system of equations in its discretized form using eq. (26), it is essential that $dA/dt \propto dF(A)/dr$, where $F(A)$ is the flux density of $A$. If $dA/dt \propto A dv/dr$ for example, some algebraic operations are needed to put $A$ inside the spatial derivative, within the routines regarding only the implicit integration scheme.

  According to this strategy, adopting ($\lambda = 0.5$, $\kappa = 1$) the continuity equation, written as:
  
\begin{equation}
\frac{d \rho}{dt} - \bmath{v} \cdot \nabla \rho + \nabla \cdot (\rho \bmath{v}) = 0
\end{equation}

in finite difference terms is converted as:
\begin{eqnarray}
\frac{3}{2} \frac{\rho_{k}^{n+1} - \rho_{k}^{n}}{\Delta t} - \frac{1}{2} \frac{\rho_{k}^{n} - \rho_{k}^{n-1}}{\Delta t} - \frac{v_{r,k}^{n+1}}{2 \Delta r_{k}} \left( \rho_{k+1}^{n+1} - \rho_{k-1}^{n+1} \right) + & & \nonumber \\ \frac{1}{2 \Delta r_{k}} \left( \rho_{k+1}^{n+1}v_{r,k+1}^{n+1} - \rho_{k-1}^{n+1} v_{r,k-1}^{n+1} \right) = 0, & &
\end{eqnarray}

where $r$ is the Cartesian component of the velocity vector. Taking into account that $k$ is a 1D grid number along each $X, Y, Z$ arbitrary directions time by time, we have $3 \times 3$ equations to be solved to get as many $\rho_{k}^{n+1}$ values to be averaged. In this last numerical expression the 1D grid number $k$, shown as a subscript, indicates the two 1D positions $k - 1$ and $k + 1$ where the physical property is SPH interpolated, being $k$ the index pointing the real particle position, along the line connecting $k -1$ to $k + 1$. In this Semi-Lagrangian explicit-implicit technique, this interpretation is applied also to any other scalar quantity or scalar vectorial component. Also the conversion of the divergence term in this numerical expression must be interpreted in like manner as an 1D conversion. This is the reason because we get $3 \times 3$ equations to be solved. Even though we use the three $X, Y, Z$ directions as preferred lines to work as $k$ directions, the choice of other arbitrary orthogonal directions does not affect the final result in so far as the adopted SPH Kernel is isotropic as it is currently used in SPH.

  In some particular cases, it is better to rewrite the same continuity equation as:

\begin{equation}
\frac{d \ln \rho}{dt} + \nabla \cdot \bmath{v} = 0.
\end{equation}

  In this case, its finite differences conversion is:

\begin{eqnarray}
\frac{3}{2} \frac{\ln \rho_{k}^{n+1} - \ln \rho_{k}^{n}}{\Delta t} - \frac{1}{2} \frac{\ln \rho_{k}^{n} - \ln \rho_{k}^{n-1}}{\Delta t} + & & \nonumber \\ \frac{1}{2 \Delta r_{k}} \left( v_{r,k+1}^{n+1} - v_{r,k-1}^{n+1} \right) = 0. & &
\end{eqnarray}

  As far as the energy equation is concerned, written as:

\begin{equation}
\frac{d \rho \epsilon}{dt} + \nabla \cdot \left[ \left(p + \rho \epsilon \right) \bmath{v} - \bmath{v} \cdot \bmath{\tau} \right] - \bmath{v} \cdot \nabla \left( p + \rho \epsilon \right) = 0
\end{equation}

we have:

\begin{eqnarray}
\frac{3}{2} \frac{\rho_{k}^{n+1} \epsilon_{k}^{n+1} - \rho_{k}^{n} \epsilon_{k}^{n}}{\Delta t} - \frac{1}{2} \frac{\rho_{k}^{n} \epsilon_{k}^{n} - \rho_{k}^{n-1} \epsilon_{k}^{n-1}}{\Delta t} + & & \nonumber \\  \frac{1}{2 \Delta r_{k}} \left[ \left( p + \rho \epsilon \right)_{k+1}^{n+1} v_{r,k+1}^{n+1} - \left( p + \rho \epsilon \right)_{k-1}^{n+1} v_{r,k-1}^{n+1} \right] - & & \nonumber \\ \frac{1}{2 \Delta r_{k}} \sum_{s} \left( \tau_{r,s,k+1}^{n+1}v_{r,k+1}^{n+1} - \tau_{r,s,k-1}^{n+1} v_{r,k-1}^{n+1} \right) - \nonumber \\ \frac{v_{r,k}^{n+1}}{2 \Delta r_{k}} \left[ \left( p + \rho \epsilon \right)_{k+1}^{n+1} - \left( p + \rho \epsilon \right)_{k-1}^{n+1} \right] = 0, & &
\end{eqnarray}

where $s = X, Y, Z$.

  As far as the momentum equation is concerned, we postpone the inclusion of all explicit external field contributions (e.g. gravitational, electric, magnetic etc.) after having computed the velocity due to the thermodynamics alone. Such contributions can be easily added to the integration as $\Delta t \nabla \Phi_{grav, k}^{n+1}$. According to the same concepts, the momentum equation, without any external field contribution

\begin{equation}
\frac{d v_{r}}{dt} + \frac{1}{\rho} \left[ \frac{d p}{dr} - \left( (\nabla \cdot \bmath{\tau}) \cdot \bmath{r} \right) \right] = 0
\end{equation}

is converted as:

\begin{eqnarray}
\frac{3}{2} \frac{v_{r,k}^{n+1} - v_{r,k}^{n}}{\Delta t} - \frac{1}{2} \frac{v_{r,k}^{n} - v_{r,k}^{n-1}}{\Delta t} + \frac{p_{k+1}^{n+1} - p_{k-1}^{n+1}}{2 \rho_{k}^{n+1} \Delta r} - & & \nonumber \\ \frac{1}{2 \rho_{k}^{n+1} \Delta r_{k}} \sum_{s} \left( \bmath{\tau}_{s,r,k+1}^{n+1} - \bmath{\tau}_{s,r,k-1}^{n+1} \right) = 0. & &
\end{eqnarray}

Instead, if the same momentum equation is written as:

\begin{equation}
\frac{d \rho \bmath{v}}{dt} - \bmath{v} \cdot \nabla (\rho \bmath{v}) + \nabla \cdot ( \rho \bmath{v} \bmath{v} - \bmath{\tau}) + \nabla p = 0,
\end{equation}

its conversion in finite difference terms is:

\begin{eqnarray}
\frac{3}{2} \frac{\rho_{k}^{n+1} v_{r,k}^{n+1} - \rho_{k}^{n} v_{r,k}^{n}}{\Delta t} - \frac{1}{2} \frac{\rho_{k}^{n} v_{r,k}^{n} - \rho_{k}^{n-1} v_{r,k}^{n-1}}{\Delta t} - & & \nonumber \\ \frac{v_{r,k}^{n+1}}{2 \Delta r_{k}} \left( \rho_{k+1}^{n+1} v_{r,k+1}^{n+1} - \rho_{k-1}^{n+1} v_{r,k-1}^{n+1} \right) + \frac{p_{k+1}^{n+1} - p_{k-1}^{n+1}}{2 \Delta r} + & & \nonumber \\ \frac{1}{2 \Delta r_{k}} \sum_{s} \left( \rho_{k+1}^{n+1} v_{r,k+1}^{n+1}  v_{s,k+1}^{n+1} - \rho_{k-1}^{n+1} v_{r,k-1}^{n+1}  v_{s,k-1}^{n+1} \right) - \nonumber \\ \frac{1}{2 \Delta r_{k}} \sum_{s} \left( \bmath{\tau}_{s,r,k+1}^{n+1} - \bmath{\tau}_{s,r,k-1}^{n+1} \right) = 0.
\end{eqnarray}

  At last, as far as the Lagrangian position updating equation $d \bmath{r}/dt = \bmath{v}$, it could be solved for each $r = X, Y, Z$ component either explicitly as:

\begin{equation}
r_{k}^{n+1} = r_{k}^{n} + v_{r,k} \Delta t
\end{equation}

or implicitly as:

\begin{equation}
\frac{3}{2} \frac{r_{k}^{n+1} - r_{k}^{n}}{\Delta t} - \frac{1}{2} \frac{r_{k}^{n} - r_{k}^{n-1}}{\Delta t}  - v_{k,r}^{n+1} = 0.
\end{equation}

  Throughout these algebraic expressions, some kinetic and thermodynamic quantities have to be computed at $(k+1), (n+1)$ and at $(k-1), (n+1)$ points in the spacetime grid. Adding the $i$ subscript, referring to the $i$th SPH particle, we get:

\begin{equation}
A_{i,k+1}^{n+1} = A_{i,k}^{n+1} + \Delta \bmath{s}_{i}^{n+1} \cdot \sum_{j=1}^{N} m_{j} \frac{A_{j,k}^{n+1}}{\rho_{j,k}^{n+1}} \nabla_{i} W_{ij}^{n+1},
\end{equation}

and

\begin{equation}
A_{i,k-1}^{n+1} = A_{i,k}^{n+1} - \Delta \bmath{s}_{i}^{n+1} \cdot \sum_{j=1}^{N} m_{j} \frac{A_{j,k}^{n+1}}{\rho_{j,k}^{n+1}} \nabla_{i} W_{ij}^{n+1},
\end{equation}

where $\Delta \bmath{s}_{i}^{n+1}$ is a vector along the direction joining the space-time interpolating grid points $(k-1), (n+1)$ to $(k+1), (n+1)$. Thus, the SPH technique is again used to compute the spatial gradients $\nabla A_{i}$ (e.g. gradients of density, pressure including its dissipation terms and so on) to the $i$th SPH particle within the scalar product in the 2nd term in eqs. (45, 46). For the sake of simplicity, even though $|\Delta \bmath{s}_{i}^{n+1}|$ is an arbitrary length, we adopt $|\Delta \bmath{s}_{i}^{n+1}| = h_{i}$, whatever is the time level $n$.

  Once $\rho_{i,s}^{n+1}$, $\epsilon_{i,s}^{n+1}$, $v_{r,i,s}^{n+1}$ ($r = X, Y, Z$) are given at time $n+1$ for the $i$th SPH particle for $s = 1, 2, 3$ in 3D, the final values $\rho_{i}^{n+1}$, $\epsilon_{i}^{n+1}$, $v_{r,i}^{n+1}$ are calculated as: $A_{i}^{n+1} = \sum_{X,Y,Z} \sum_{s}  A_{i,s}^{n+1}/9$ where $A_{i,s}^{n+1}$ , where the two summations take into account of the three flow components for each of the three directions ($3 \times 3$).

  In this notation, the $s$ index refers to the direction of the 1D arbitrary computational spatial line, while the $r$ subscript refers to vectorial components. In our notation, $s$ follows $r$, but it could be different, in principle. Therefore, as an example, we compute $\rho_{i,x}^{n+1}$, $\rho_{i,y}^{n+1}$, $\rho_{i,z}^{n+1}$, $\epsilon_{i,x}^{n+1}$, $\epsilon_{i,y}^{n+1}$, $\epsilon_{i,z}^{n+1}$ that means the density and the thermal energy per unit mass on the ith particle, at time $n+1$, along the $x, y, z$ lines (along the $1, 2, 3$ arbitrary lines). Moreover, $v_{x,i,x}^{n+1}$, $v_{x,i,y}^{n+1}$, $v_{x,i,z}^{n+1}$, $v_{y,i,x}^{n+1}$, $v_{y,i,y}^{n+1}$, $v_{y,i,z}^{n+1}$, $v_{z,i,x}^{n+1}$, $v_{z,i,y}^{n+1}$, $v_{z,i,z}^{n+1}$ mean the $r = x, y, z$ velocity components on the ith particle, at time $n+1$, along the $s = x, y, z$ lines (along the $1, 2, 3$ arbitrary lines).

  The strategy adopted in this explicit-implicit Semi-Lagrangian scheme is similar, for some aspects to that adopted to solve ordinary differential equations (ODE), even up to the second order in some explicit and implicit three point methods by some authors \citep{c28,c29}.

  Errors in a Semi-Lagrangian explicit-implicit integration method include errors of the backward implicit integration ($O(\Delta t)$), and errors from the interpolation: $E(\Delta x) = O(\Delta x_{k-1}, \Delta x_{k+1})$. Therefore, the overall accuracy of the method is

\begin{equation}
\frac{u_{k}^{n+1} - u_{k}^{n}}{\Delta t} = \frac{du}{dt} + O\left( \Delta t + \frac{O(\Delta x_{k+1}^{n+1}, \Delta x_{k-1}^{n+1})}{\Delta t} \right).
\end{equation}

  An exact calculation of (47) can be found in \citep{b12}. Eq. 47 shows that the error is not monotonic with respect to $\Delta t$. Errors coming from a poor spatial interpolation could dominate. As $\Delta t $ increases, the overall error decreases. If the first term $O(\Delta t)$ is insignificant, a further increase of $n$ will not improve the overall accuracy. For the sake of simplicity, we consider negligible errors in the SPH spatial interpolation at the time step level $n+1$. This relevant argument is beyond the scope of this paper. On the other hand, if SPH spatial interpolation is inadequate, even the explicit integrations do not improve the quality of results. Hence, if the solution is well-resolved in space with efficient spatial interpolations, the overall consistency (accuracy + stability + convergence) is solely due to the implicit integration method that, for a 3FICS is well documented in various textbooks (for example \citet{c3,c37,c5,c4}).

\subsection{Choice of a time step for an SPH Semi-Lagrangian explicit-implicit scheme}

  A time step restriction is always necessary for time dependent calculations in computational fluid dynamics. Restrictions are needed for mathematical stability in explicit calculations. Instead they are necessary for accuracy considerations in implicit calculations. The Courant-Friedrichs-Lewy condition on the time step progression to solve PDE and ODE for explicit integration techniques offers a temporal reference where numerical solutions are both stable and convergent with the mathematical solutions. Unfortunately, such a temporal reference is still debated for implicit integration of PDE and ODE \citep{c23,c20}.

  For SPH technique, the explicit time limiter is given by:
  
\begin{equation}
\Delta t_{SPH} = C {\textrm min}_{i=1,N} \left[ \frac{h}{v_{sig,ij}}, |\nabla \cdot \bmath{v}_{i}|^{-1}, \left( \frac{h}{|\bmath{a}_{i}|} \right)^{1/2} \right],
\end{equation}

which includes the Courant-Friedrichs-Lewy time limiter $\Delta t_{CFL}$. $v_{sig,ij}$ is the signal transmission velocity between close particles $i$ and $j$ within the SPH spatial resolution length $h$ \citep{c11,c12,c39,c54}, also including the sound velocity $c_{si}$, while $|\bmath{a}|_{i}$ is the full acceleration for the $i$th SPH particle. $C$ is a number of the order of $0.2 - 0.5$.

  Since in a correct Free Lagrangian particle fluid dynamics particles cannot interpenetrate with each other, another time step limit is the "kinetic" value:

\begin{equation}
\Delta t_{k} = C {\textrm min}_{i=1,N} \left[ |\nabla \cdot \bmath{v}_{i}|^{-1}, \left( \frac{h}{|\bmath{f}_{i}|} \right)^{1/2} \right],
\end{equation}

where $|\bmath{f}|_{i}|$ is the force per unit mass only due to gravitational and to pressure terms. However, such a longer limiter, could be correct for a Semi-Lagrangian implicit SPH hydrodynamics only for weak shock flows. To overcome such a difficulty, we consider the limiter:

\begin{equation}
\Delta t_{l} = \left( \Delta t_{SPH} \Delta t_{k} \right)^{1/2},
\end{equation}

which corresponds to the geometric mean of the previous two. This choice both largely prevents the numerical collapse of the time step to be used, and gives a good hydrodynamics far from the risk of getting wrong solution for supersonic flows that occur when $\Delta t \sim \Delta t_{k}$. Of course, for very extreme situations, even this time step limiter cannot be correctly used, because of the absence of a temporal top limit for implicit integrations techniques. The ratio $\Delta t_{k}/\Delta t_{SPH}$ is always $\geq 1$. However, whenever $\Delta t_{k}/\Delta t_{SPH} > 400$ (e.g. $\Delta t_{l} > 20 \Delta t_{SPH}$), some consistent deviations from the correct solutions are recorded. These deviations are much better confined in so far as $\Delta t_{k}/\Delta t_{SPH} \leq 10$.

  Hence, the choice of a time step for a Semi-Lagrangian explicit-implicit or for an implicit scheme is totally arbitrary, in principle, because it does not exist a theory predicting a time step limiter. For methods using implicit integration, it could also be possible to adopt $t^{n+1} - t^{n} = \Delta t \propto \Delta t_{SPH}$, whose parameter of proportionality should be found case by case. Formulation (50) automatically simply correlates the implicit $\Delta t$ to $\Delta t_{SPH}$ and to $\Delta t_{k}$. It is true that sometimes we also could implicitly work using a larger $\Delta t$, but some cautions are needed, because accuracy of solution could surely be compromised.

  Given these definitions, the time step adopted for the Semi-Lagrangian explicit-implicit scheme discussed in \S 3.2 is: $t^{n+1} - t^{n} = \Delta t = \Delta t_{l} - \Delta t_{SPH}$, being $\Delta t_{SPH}$ used in the first explicit step. Whenever $\Delta t = 0$, only the explicit technique is applied.

  Throughout the rest of the paper, we adopted $C = 0.25$ in the evaluation of $\Delta t_{SPH}$, $\Delta t_{k}$. A larger $C$ value is always possible, up to $C = 0.5$ without relevant instabilities in the explicit solution. This implies that the longer $\Delta t_{SPH}$ will involve longer $\Delta t_{l}$, according to the choice (50).

\section{Critical tests}

  What is previously discussed shows that the system of equations of hydrodynamics in its Lagrangian form (eqs. 1 - 3) should be algebraically manipulated in a form similar to the finite difference methods, conserving the Lagrangian formalism.

  To verify whether the Semi-Lagrangian explicit-implicit approach is correct, discontinuity in the flow represents critical tests especially at the discontinuity front. The case of flat translational flows, or the propagation of linear sound waves, does not represent critical tests. In these cases, either trivial simplifications of the non-linear Euler system of equation, or negligible discrepancies in the flow within SPH spatial resolution length $h$, detectable on scale length longer than  $h$, do not produce sensible differences in the terms included in the continuity, energy and momentum equations (\S 3.2). To this purpose, we present both a SPH 1D and a 2D tests in the case of a blast wave and a SPH 2D test regarding a 2D shockless radial viscous transport in an annulus ring. The 2D modelling is also interesting to check the correctness and the consistency of the arbitrary statistical criterion of producing the integrated numerical solution by the average of as many integrals as many are the dimensions of the flow (\S 3.2). At the same time, these simulations allow us to verify whether the choice for the time step (\S 3.3) is free of criticisms. What is important is that SPH explicit and SPH Semi-Lagrangian explicit-implicit results of integration strongly compare with each other.

\begin{figure*}
\resizebox{\hsize}{!}{\includegraphics[clip=true]{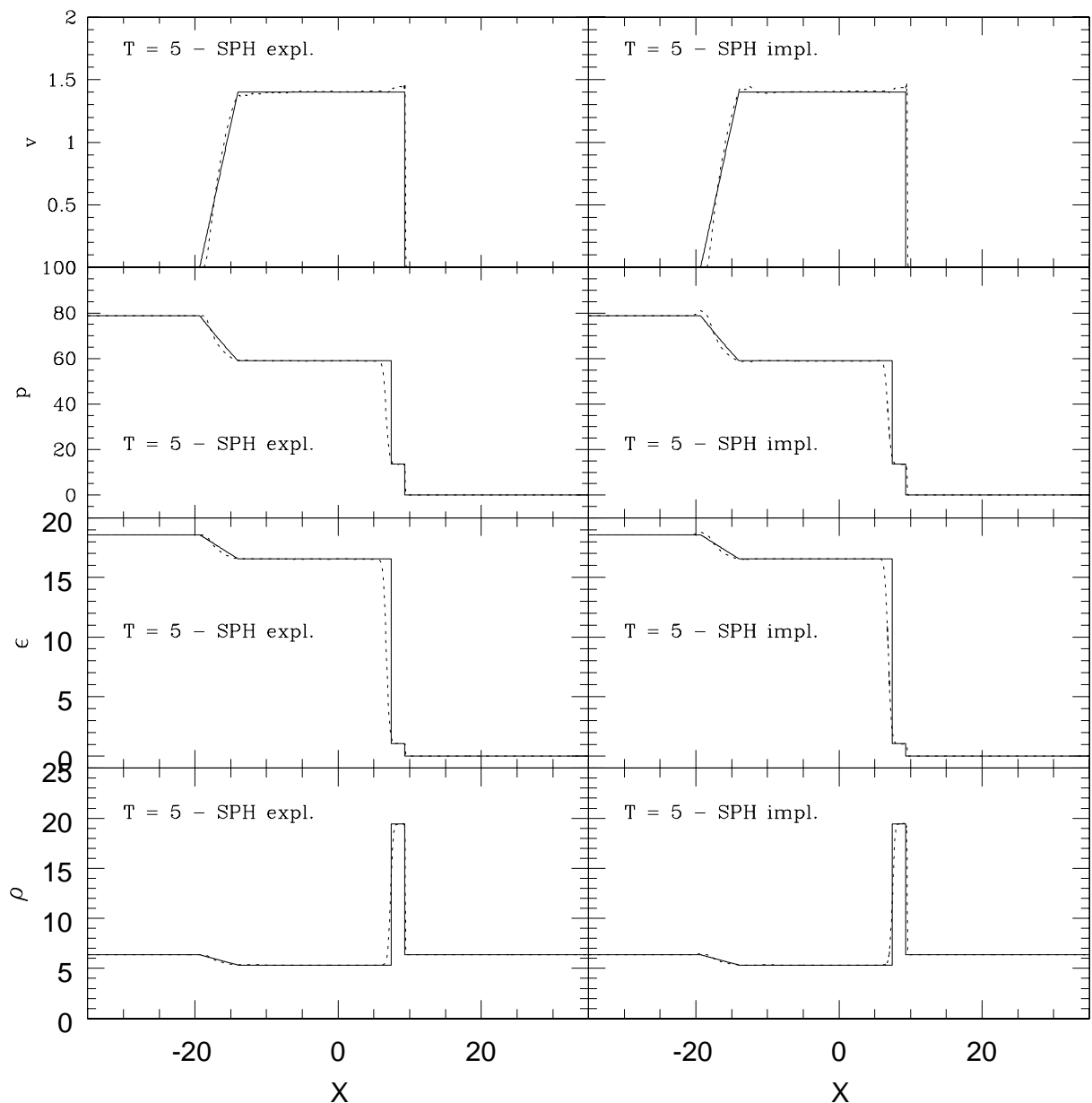}}
\caption{1D blast wave test regarding the comparison of both explicit (left size) and implicit (right size) SPH results to analytical (solid line) results. Density $\rho$, thermal energy $\epsilon$, pressure $p$ and velocity $v$ are plotted at time $T = 5$. Density and thermal energy of particles initially at rest at time $T = 0$ refer to values plotted at the two edges for each plot. The initial velocity is zero throughout.}
\end{figure*}

  Even though a variable smoothing resolution length could also be used in SPH codes, the inadequacy of this approach in particular as for the radial transport whenever free boundary edges characterize the flow, recently discussed \citep{c58,c25}, suggests us in using a constant $h$. However, the adoption of an implicit integration scheme to SPH is not correlated to its resolution length variability or constancy.

  Being the Semi-Lagrangian explicit-implicit integration involving a more laborious algorithm spending $\approx 10\%$ more time for each iteration, compared with the simpler explicit integration algorithm, there are not practical advantages in using the Semi-Lagrangian implicit integration mechanism in so far as $\Delta t_{l} \approx 1.4 - 1.5 \Delta t_{SPH}$. However, even examples where the time step gain ratio is restricted only to $1 < \Delta t_{l}/\Delta t_{SPH} \leq 2$ are useful to understand the good and the bad of the implicit algorithm. Technically $1 < \Delta t_{l}/\Delta t_{SPH} \leq 6$ \citep{c21,c22} to avoid a relevant degradation in the accuracy of solution. This means that in so far $1 < \Delta t_{l}/\Delta t_{SPH} \leq 6$ is respected, tests are meaningful. Otherwise some unexplored side of numerical techniques is faced. For this reason, tests here proposed are not so stressed out for working with a ratio of time steps beyond the limit of $6$. A risk to be faced for computational research scopes is the adoption of a Semi-Lagrangian approach even if $\Delta t_{l}/\Delta t_{SPH} \approx 5 - 10$, betting that the accuracy in the solutions is quite good and that the larger numerical dissipation is limited \citep{c23}. In this case, the reduction of time in calculations is $\approx 2 - 3$ times compared with the explicit calculations. However, sometimes, the gain-effectiveness of the Semi-Lagrangian technique depends on the context of the fluid dynamics problem \citep{b16}, as well as on the spatial discretization \citep{b12}.

\subsection{1D and 2D blast wave tests}

  The behaviour of shock fronts moving in the prevailing flow is analytically described by the Rankine-Hugoniot "jump conditions" \citep{c37,c5,c4,c38,c36}. These conditions are obtained by spatially integrating the hyperbolic Euler equations across the discontinuity between the two flow regimes left-right ($l - r$) in their Eulerian formalism. In 1D, such equations are:

\begin{eqnarray}
\frac{\partial \rho}{\partial t} & = & - \frac{\partial}{\partial x} (\rho v) \\
\frac{\partial \rho v}{\partial t} & = & - \frac{\partial}{\partial x} (\rho v^{2} + p) \\
\frac{\partial \rho E}{\partial t} & = & - \frac{\partial}{\partial x} [\rho v (E + p/\rho)],
\end{eqnarray}

where $E = v^{2}/2 + \epsilon$, whose conservative analytical form can be synthesized as:

\begin{equation}
\frac{\partial w}{\partial t} = - \frac{\partial}{\partial x} f(w).
\end{equation}

  Whenever in a shocktube the ratios $p_{l}/p_{r} = \epsilon_{l}/\epsilon_{r} \gg 1$ (and consequently $\rho_{1}/\rho_{2} = 1$), and $v_{l} = v_{r} = 0$, such a discontinuity is called a "blast wave". Fig. 2 shows a comparison of both explicit and implicit SPH results together with the so called analytical solution. In the SPH blast wave test here considered, $p_{1}/p_{1} = \epsilon_{1}/\epsilon_{2} = 10^{4}$, while the SPH particle resolution length is $h = 5 \cdot 10^{-2}$. The whole computational domain is built up with $2001$ particles from $X = 0$ to $X = 100$, whose mass is different, according to the shock initial position. At time $T = 0$ all particles are motionless and the adiabatic index $\gamma = 5/3$, while the ratios $\rho_{1}/\rho_{2} = 1$. The first $5$ and the last $5$ particles of the 1D computational domain, keep fixed positions and do not move. The choice of the final computational time is totally arbitrary, since the shock progresses in time. Initial values of $v$, $\rho$ and $\epsilon$ at time $T = 0$ are shown at the left-right edges of each plot of the same Fig. 2.

  Fig. 2 shows that both SPH results globally compare with each other and that they also compare with the analytical solution. Both SPH (explicit and implicit) results are in a good comparison with the analytical solution. Discrepancies involve only $4$ particles at most, with the exception of numerical solutions corresponding to analytical vertical profiles regarding thermal energy where, for both SPH solutions, discrepancies are larger. As Fig. 2 clearly shows, both the SPH numerical solutions suffer from some small well known instabilities, especially in the proximity of discontinuities \citep{c40,c41,c39} as far as the velocity profile is concerned. Such effect comes out whenever a spatial high resolution is working together with an explicit handling of dissipation through an artificial viscosity damping to solve the Riemann problem of flow discontinuities. A low spatial resolution hides this effect because of the higher artificial damping due to a higher particle resolution length $h$ (eqs. 17-18). The higher the spatial resolution (the smaller $h$), the higher the "blimp" instabilities. Moreover, in SPH, even the choice of the arbitrary parameters $\alpha_{SPH}$ and $\beta_{SPH}$ should be linked to the specific physical problem. It is true that it is possible to slightly modify the artificial viscosity strength if only weak shocks appear in the problem at hand, thus, in general it is neither desirable nor necessary to tune the $\alpha_{SPH}$ and $\beta_{SPH}$ parameters to a specific problem. However, even $\alpha_{SPH} = 0.04$ and $\beta_{SPH} = 0$ were also used \citep{c19} for turbulence development. Another way for reduction of instabilities is possible \citep{c16,c24,c25} if the damping is strictly locally physical, using eqs. (23, 24) instead of eq. (19) for the perfect gases EoS. An effective reduction of "blimp" effects is obtained, especially for 1D blast waves, where strong discontinuities in the flow deeply affect the SPH numerical solution producing intrinsic numerical instabilities close to the propagating discontinuities.

\begin{figure}
\resizebox{\hsize}{!}{\includegraphics[clip=true]{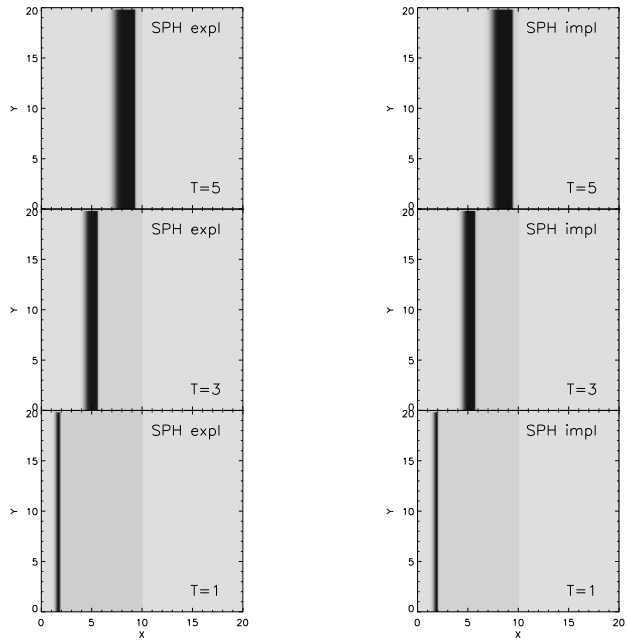}}
\caption{Examples of density wave propagation for the 2D blast wave tests. The density wave front develops from $X = 0$ at time $T = 0$. Only plots at times $T = 1, 3, 5$ are represented. Density isocontour maps in 64 greytones are shown in each side.}
\end{figure}

\begin{figure}
\resizebox{\hsize}{!}{\includegraphics[clip=true]{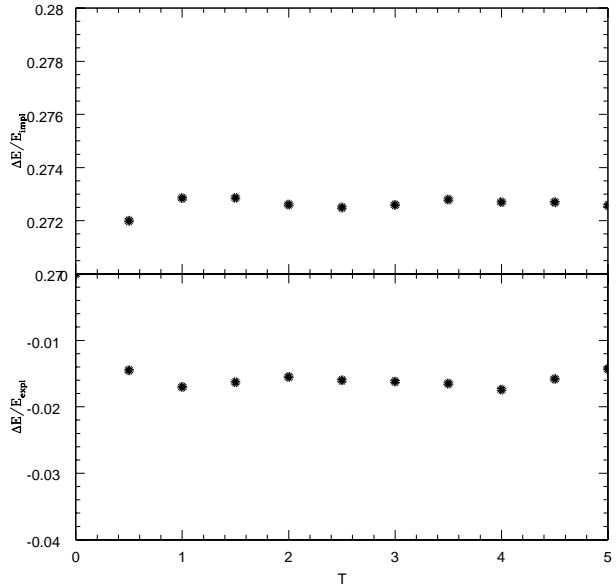}}
\caption{$\Delta t_{l}/\Delta t_{SPH}$ and relative errors in the total energy $\Delta E/E$ per unit mass, from the total energy $E$ conservation, reported as a function of time $T$ for the 2D blast wave for both explicit and explicit-implicit integrations. Notice that for the sake of simplicity of representation, these errors are multiplied by $10^{3}.$. }
\end{figure}

  To further check the consistency and the validity of the Semi-Lagrangian explicit-implicit integration method, we also performed both explicit and explicit-implicit 2D blast wave tests on the basis of the same initial physical parameters, considering an initial ordered disposition of particles, adopting the same particle resolution length $h$ and the same $\Delta x = h = 0.05$ and $\Delta y = h = 0.05$ particle separation. The four edges of the box at $x_{min} = - 20$, $x_{max} = 20$, $y_{min} = 0$, $y_{max} = 20$ are formed of three equally spaced lines of always motionless particles. As in the previous example, the blast wave starts from $x = 0$. The temporal progress of the density wave is shown in Fig. 3 for both the integration techniques. For the sake of simplicity, only greytone contours are shown at time $T = 1, 3$ and $5$. The agreement looks like quite good. However, as we discussed in \S 3.1, despite stable, the implicit integration techniques involves greater errors compared with those made in explicit integrations. Fig. 4 shows the relative error for both integration schemes on the total energy $\Delta E/E$, where it is clearly shown that in the Semi-Lagrangian explicit-implicit integration scheme, relative errors on the total energy per unit mass per particle $E = v^{2}/2 + \epsilon$ are larger, being of the order of $\sim 2 \cdot 10^{-4}$ against $\sim 2 \cdot 10^{-5}$, relative to the explicit integration scheme, as can be evaluated by the two pictures shown in this figure. However, what is relevant is that in the Semi-Lagrangian explicit-implicit integration method such error does not increase in time compared with that relative to the explicit integration technique. The most of the relative error is made at the first stages of integrations in both cases, after that it looks like to be stationary.

  Apart from the accordance between analytical and numerical (explicit and implicit) results, what is remarkable is the fact that implicit results are obtained in about half of cpu time in implicit SPH approach, compared with the traditional explicit SPH, adopting the time step criterion expressed by eq. (46). The final time configuration at time $T = 5$ is an evolved configuration well beyond the initial instants from $T = 0$. At the final time configuration at time $T = 5$ the time step gain ratio $\Delta t_{l}/\Delta t_{SPH} \sim 1.6$. Since the beginning of a blast wave simulation, a time step gain ratio $\sim 2$ is quickly achieved as an order of magnitude. At the beginning of simulations, the time stepping is principally governed by the thermal contribution thanks to the sudden increase in the velocity. Time by time, the decrease of $v$ and the relative decrease in $\Delta t_{k}$, set a larger and larger $\Delta t_{l}$ (derived by $\Delta t_{k}$) with respect to $\Delta t_{SPH}$. This implies that in the particular case of a detonation, the criterion adopted on the time step progress ($\Delta t_{l}$ in \S 3.3) shows accordance with analytical results without any difficulty, being $\Delta t_{l}/\Delta t_{SPH}$, well within the factor of $6$ discussed in \S 1. Notice that this accordance is not verified at the 1st time step, because an explicit integration scheme must be applied.

\subsection{2D shockless isothermal radial viscous transport in an annulus ring}

  Theory on 2D shockless radial transport in a Keplerian annulus ring in a gravitational potential well \citep{b3} predicts that, from the Green function, the solution of the initial Keplerian mass distribution at time $t = 0$ for $\Sigma$ is:

\begin{equation}
\Sigma(r,t=0) = m \delta(r - r_{\circ})/2\pi r_{\circ}
\end{equation}

for a ring of mass $m$, with an initial radius $r_{\circ}$, The solution, at time $t$, in terms of dimensionless radius $x = r/r_{\circ}$ and viscous time $\theta = 12 \nu t r_{\circ}^{-2}$ is

\begin{equation}
\Sigma(x,t) = (m/\pi r_{\circ}^{-2}) \theta^{-1} x^{-1/4} \exp[-(1 + x^{2})/\theta] I_{1/4}(2x/\theta),
\end{equation}

\begin{figure}
\vspace{-4.3 cm}
\resizebox{\hsize}{!}{\includegraphics[clip=true]{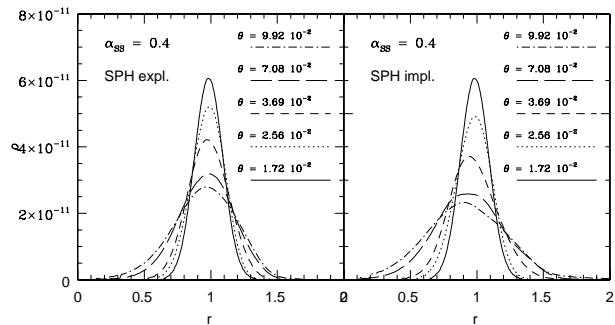}}
\caption{Comparison of radial density distributions of a viscous annulus ring for both explicit (left size) and implicit (right size) SPH results in the Shakura and Sunyaev formulation using $\alpha_{SS} = 0.4$. Viscous time $\theta$ for each distribution is also reported.}
\end{figure}

where $I_{1/4}$ is the modified Bessel function. The action of viscosity is to spread out the entire annulus ring toward a disc structure transporting most of the low angular momentum mass toward the centre of the potential well and transporting a smaller fraction of high angular momentum mass toward the empty external space.

  For practical computational purposes, since it is impossible to reproduce a delta Dirac function at time $t = 0$ for the initial mass distribution, it is necessary to start numerical calculations from an initial mass distribution relative to a small $\theta$ value. In our example, we choose $\theta = 0.017$, a value comparable to that used by \citet{b11} for a radial viscous transport similar test.

\begin{figure}
\resizebox{\hsize}{!}{\includegraphics[clip=true]{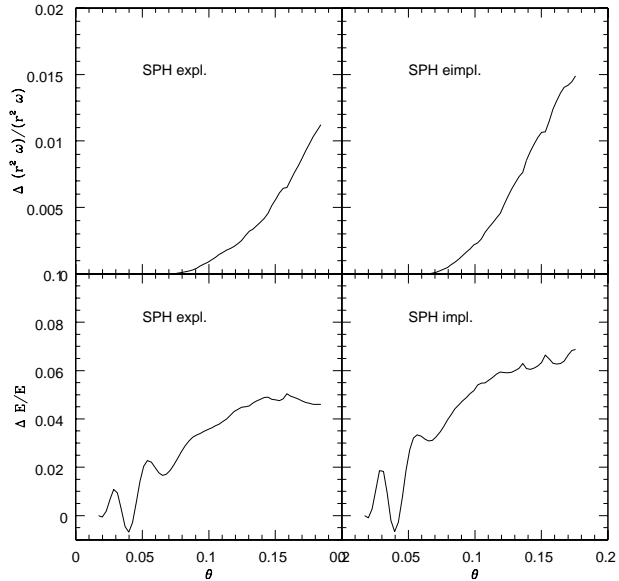}}
\caption{Comparison of relative errors for the total energy $E$ per unit mass and for the specific angular momentum $r^{2} \omega$, for the radial viscous transport in an annulus ring as a function of the viscous time $\theta$. Explicit and Semi-Lagrangian explicit-implicit relative errors are compared.}
\end{figure}

\begin{figure}
\resizebox{\hsize}{!}{\includegraphics[clip=true]{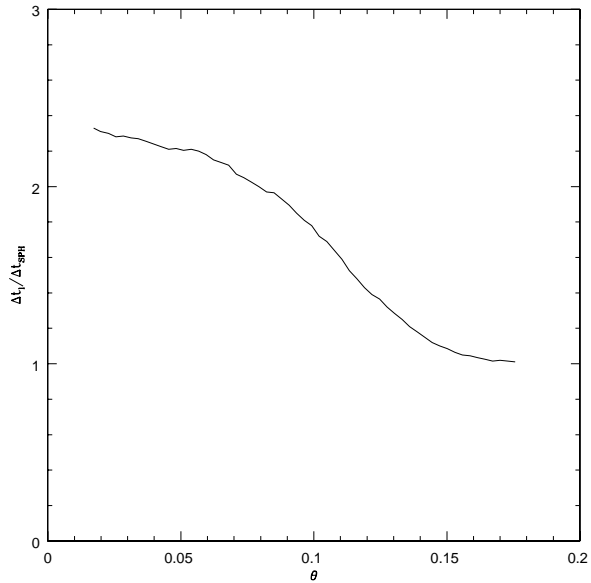}}
\caption{Implicit to explicit time step $\Delta t_{l}/\Delta t_{SPH}$ reported as a function of the viscous time $\theta$ for the viscous spreading of the annulus ring.}
\end{figure}

  The initial setting is that of a 2D annulus ring including $40000$ particles whose $h = 0.09$ forming a ring whose radius $r_{\circ} = 1$ around a star having mass $M = 1$. The initial radial spread at $\theta = 0.017$ is $\Delta r = 1$. Throughout the simulations $d \epsilon /dt = 0$, that means that the entire configurations is permanently isothermal. $\gamma = 5/3$ and the unifirm sound velocity $c_{s} = 5$ to have low Mach numbers from the beginning of the simulations. In reality, a value of $c_{s} \simeq 0.05$, like that adopted by \citet{b11} would not address any time stepping advantage to the implicit integrations because $\Delta t_{l} \simeq \Delta t_{SPH}$ in isothermal conditions, preventing any flow heating. For a dimensional arbitrary initial setting, we consider a central star having $M_{\circ} = 1 M_{\odot}$, while the dimensional annulus ring radius $R_{\circ} = 10^{11}$ cm. These are also normalization factors for masses and lengths for a dimensionless handling of equations, as previously discussed in \S 3.2. $v_{\circ} = 2 \pi (GM_{\circ}/R_{\circ})^{1/2}$ is the normalization factor for velocities, so that the orbital period is normalized to $1$.

  Figure 5 compares results on the radial distribution of the 2D mass density $\Sigma$ for the viscous annulus ring at various viscous times $\theta$ both for SPH explicit integrations and for Semi-Lagrangian explicit-implicit integrations for a \citet{c50,c51} viscosity parameter $\alpha_{SS} = 0.4$. In figure 5, these two pictures clearly show that, $\Sigma(r,\theta)$ for implicit integration suffers of the cumulation of both explicit and iterated implicit dissipation viscosity due to the effect of a further intrinsic viscosity algebraically developed by repeated implicit iterations. Even though fortunately limited, relative errors in figure 6 for the implicit integration scheme on the total energy per unit mass $E = v^{2}/2 + \epsilon - 1/r$ and on the specific angular momentum, as a function of viscous time $\theta$, are moderately larger compared with those relative to the explicit integration scheme. This is not a surprise, since it is previously discussed why errors in the implicit integration techniques are larger. The moderately larger inaccuracy is however compensated by a larger stability in results and by a shorter computational time. In the example, here discussed, we yielded implicit results after a total time $\sim 1/2$ shorter than the total time to conclude explicit calculations. In Fig. 7, it is shown the time step gain ratio $\Delta t_{l}/\Delta t_{SPH}$ as a function of the viscous time $\theta$. At the beginnning of the simulation the explicit $\Delta t_{SPH}$ is substantially due to the thermal contribution, so that the ratio between the two time steps is at its maximum, being the tangential kinematics substantially Keplerian and being the annulus ring still at its initial configuration. Once time progresses, $\Delta t_{k}$ becomes shorter as a consequence of the inner edge Keplerian kinematics, so that differences between the explicit and the implicit time stepping reduce. In this example, the decrease of the kinematic $\Delta t_{k}$ time by time caused by the inner edge contraction toward shorter $r$, reduces differences between the thermal explicit $\Delta t_{SPH}$ to the implicit $\Delta t_{l}$.

  Results of this test clearly show the good and the bad side of a Semi-Lagrangian explicit-implicit integration techique compared with a simpler explicit integration technique in the worst situation when in the case of isothermal flows, there is not any advantage in using an implicit time stepping. Solution are stable and accurate, in so far as $\Delta t_{l}/\Delta t_{SPH}$ is not too large as previously discussed, but results are affected by a larger numerical dissipation developed by the iterative implicit integration procedure. It is this larger dissipation that ensures a better stability of solutions at the price of a moderate inaccuracy of solution. Despite this example shows how the implicit integration is not competitive to the explicit technique in these situations, results here shown illustrate how much is the difference in explicit and implicit results in the worst competitivity conditions to the implicit integration scheme. The presence of collisional shocks, here prevented, would decrease $\Delta t_{SPH}$, better stressing the difference between $\Delta t_{SPH}$ and $\Delta t_{l}$.

\section{Simulations of a 3D accretion disc around a MBH in a close binary}

  We compare results of both inviscid and viscous stationary disc structures performing SPH simulations whose integration schemes are both explicit and implicit in low compressibility ($\gamma = 1.3$) with the aim of getting a physically well-bound accretion disc around a MBH a close binary. Previous preliminary results on this theme were published \citep{c42,c43,c44} both in 2D and in 3D. 

  The characteristics of the binary system are determined by the masses of the MBH and of its companion star and their separation. We chose to model a system in which the mass $M_{1}$ of the primary MBH and the mass $M_{2}$ of the secondary subgiant star are equal to $32 M_{\odot}$ and $1 M_{\odot}$, respectively and their mutual separation is $d_{12} = 10^{8} \ Km$. The primary's potential well is totally empty at the beginning of each simulation at time $T = 0$. The injection gas velocity at L1 is fixed to $v_{inj} \simeq 130 \ Km \ s^{-1}$ while the injection gas temperature at L1 is fixed to $T_{\circ} = 10^{4} \ K$, taking into account, as a first approximation, the radiative heating of the secondary surface due to radiation coming from the disc. Gas compressibility is fixed by the adiabatic index $\gamma = 1.3$. Supersonic kinematic conditions at L1 are discussed in \citet{c45,c46,c47}, especially when active phases of CB's are considered. However, results of this paper are to be considered as a useful test to check whether disc structures (viscous and non) show the expected behaviour. The reference frame is centred on the primary compact star, whose rotational period, normalized to $2 \pi$, coincides with the orbital period of the binary system. This explains why in the momentum equation (eq. 2), we also include the Coriolis and the centrifugal accelerations.

  Pressure, density, temperature and velocity are six unknowns to be found. Therefore we solve the continuity, momentum, energy, and EoS equations. In order to make our equations dimensionless, we adopt the following normalization factors: $M = M_{1} + M_{2}$ for masses, $d_{12} = 10^{11} \ cm$ for lengths, $v_{\circ} = (G(M_{1} + M_{2})/d_{12})^{1/2}$ for speeds, so that the orbital period is normalized to $2 \pi$, $\rho_{\circ} = 10^{-9} \ g \ cm^{-3}$ for the density, $p_{\circ} = \rho_{\circ} v_{\circ}^{2} \ dyn \ cm^{-2}$ for pressure, $v_{\circ}^{2}$ for thermal energy per unit mass and $T_{\circ} = (\gamma -1) v_{\circ}^{2} \ m_{p} \ K_{B}^{-1}$ for temperature, where $m_{p}$ is the proton mass and $K_{B}$ is the Boltzman constant. The adopted Kernel smoothing resolution length is $h = 5 \cdot 10^{-3}$ throughout. The geometric domain, including disc particles, is a sphere of radius $1$, centred on the primary MBH. The rotating reference frame is centred on the compact primary and its rotational period equals the orbital one. We simulated the physical conditions at the inner and at the outer edges as follows:

\begin{figure}
\resizebox{\hsize}{!}{\includegraphics[clip=true]{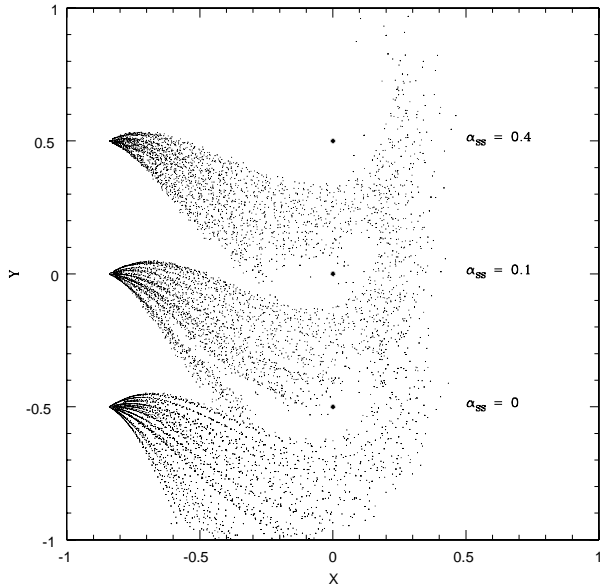}}
\caption{XY plots of the injected particle stream for the non viscous ($\alpha_{SS} = 0$) and the two disc models ($\alpha_{SS} = 0.1$ and $\alpha_{SS} = 0.4$) at time $t = 1$. The central MBH position is also shown. Notice that the $Y$ positions are shifted for $\alpha_{SS} = 0$ and for $\alpha_{SS} = 0.4$.}
\end{figure}

a) inner edge: \\
the free inflow condition is realized by eliminating particles flowing inside the sphere of radius $10^{-2}$, centred on the MBH. Although disc structure and dynamics are altered near the inner edge, these alterations are relatively small because they are balanced by a high particle concentration close to the inner edge in supersonic injection models.

b) outer edge: \\

the injection of "new" particles from L1 toward the interior of the primary Roche Lobe is simulated by generating them in fixed points, called "injectors", symmetrically placed within an angle having L1 as a vertex and an aperture of $\sim 57^{\circ}$. Normally, as adopted since our first paper on SPH accretion disc in CB \citep{c48}, the radial elongation of the whole group of injectors is $\sim 10 h$. The initial injection particle velocity is radial with respect to L1. In order to simulate a constant and smooth gas injection, a "new" particle is generated in the injectors whenever "old" particles leave an injector free, inside a small sphere with radius $h$, centred on the injector itself. Particle masses are determined by the assumed local density at the inner Lagrangian point L1: $\rho_{L1} = 10^{-9} g \ cm^{-3}$ (as typical stellar atmospheric value for the secondary star), equal to $m = \rho_{L1} (h d_{12})^{3}/(M_{1} + M_{2})$.

\begin{figure}
\resizebox{\hsize}{!}{\includegraphics[clip=true]{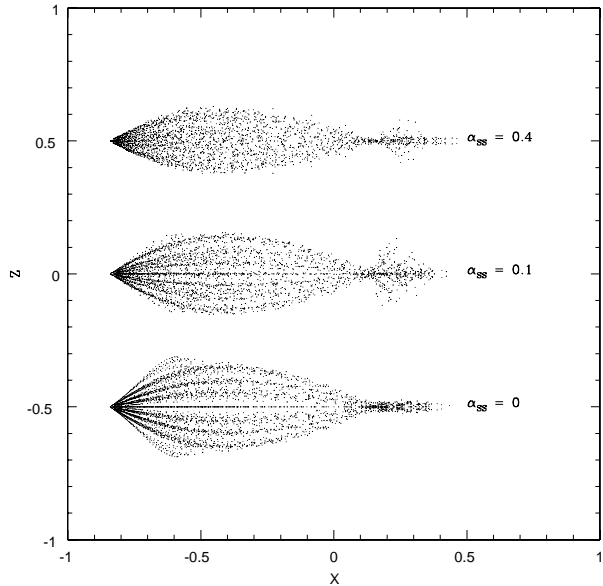}}
\caption{XZ plots of the injected particle stream as in Fig. 8.}
\end{figure}

  The adoption of supersonic mass transfer conditions from L1 is fully discussed in \citet{c45,c46}, where disc instabilities, responsible for disc active phases of CB are discussed in the light of local thermodynamics. Whenever a relevant discrepancy exists in the mass density across the inner Lagrangian point L1 between the two stellar Roche lobes, a supersonic mass transfer occurs as a consequence of the momentum flux conservation. The same result can also be obtained \citep{c49} by considering either the restricted problem of three bodies in terms of the Jacobi constant or the Bernoulli's theorem. Moreover, and this is the most important thing, we need to compare 3D disc models where $\Delta t_{SPH}$ and $\Delta t_{k}$ could be significantly different. This condition is searched for since the injection conditions at L1, favouring violent collisions among low compressibility ($\gamma = 1.3$) particles moving around a MBH, rising up particle heating at expense of the kinetic + gravitational energies. These conditions make much smaller the thermal contribution in $\Delta t_{SPH}$ evaluation compared with the kinetic contribution throughout the disc, especially when viscous heating or other forms of heating are considered. Thus when $\Delta t_{l}/\Delta t_{SPH} \gg 1$, a significant deviation in the implicit solutions would affect the whole result, up to compromising the numerical stability. A sensible reduction in $\Delta t_{SPH}$ even happens whenever other forms of signal transmission velocity (e.g. the Alf\'en speed) are also taken into account. This makes the inner disc region, at its inner edge, quite critical for a low compressibility highly viscous accretion disc around a MBH, because all these characteristics contribute to a sharp decrease of the explicit $\Delta t_{SPH}$. In particular a supersonic mass viscous flow transfer at L$_{1}$ and a central MBH determine an initial mechanical energy large eonugh to be converted into heat with high $\gamma$ to yield very low Mach numbers at the disc inner edge.

  The first viscosity coefficient (eq. 10) in the viscous disc models is related to the Shakura and Sunyaev parametrization \citep{c50,c51} as: $\eta_{v} = \rho \alpha_{SS} c_{s} H$, being $H$ the local disc thickness. The second viscosity coefficient, related to the bulk viscosity is not taken into account for the sake of simplicity. Two viscous disc models are considered, whose $\alpha_{SS} = 0.1$ and whose $\alpha_{SS} = 0.4$. These values are in accordance with the typical and with the maximum $\alpha_{SS}$ compatible with both astrophysical observations \citep{c52} and with laboratory experiments \citep{c53}. These disc models are also compared with the non viscous disc model counterpart, taken as a reference model.

  Figg. 8 and 9 show three XY and XZ plots of the injected stream from L1 at time $t = 1$ for the three disc models. The greater compactness of the injected flow of particles is visible with the increase of the viscosity parameter as a result of the sticking viscous effect. The large stream geometric spread is mainly due to the low compressibility here adopted ($\gamma = 1.3$), while the wide initial circularization radius is explained by the high angular momentum at L1 of the injected flow. These initial kinematic conditions affect the whole disc structure and kinematics throughout the simulations. Indeed, the injected gas stream yields an impenetrable boundary of the outer disc edge itself and a significant fraction of the disc's ejection flow comes from this side of the outer disc edge. Besides, a wide spray of direct cold injected flow above and below the mean disc plane also plays a role.

\begin{figure}
\resizebox{\hsize}{!}{\includegraphics[clip=true]{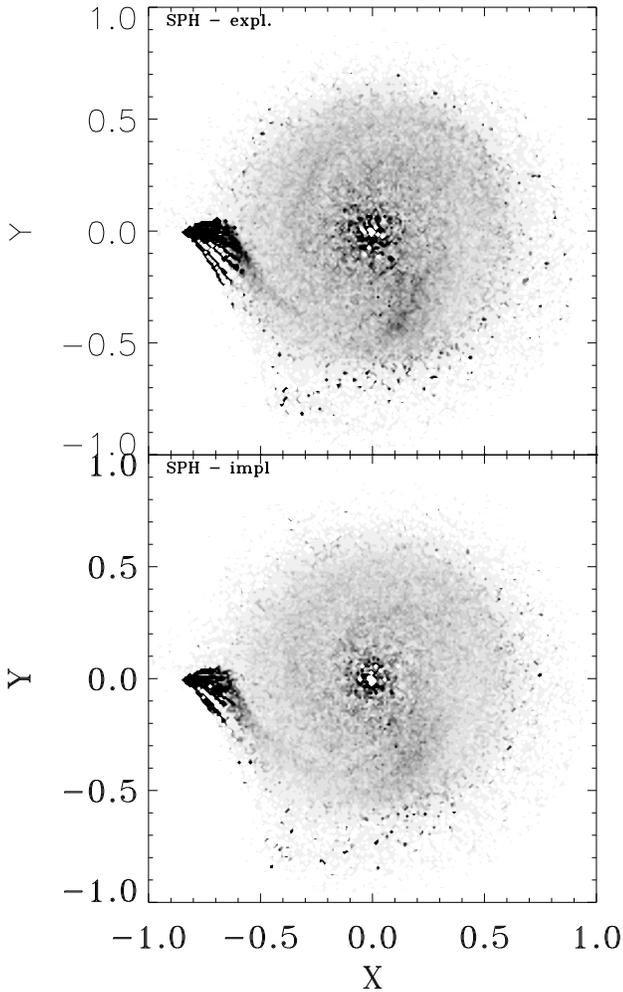}}
\caption{XY plots of $64$ greytones density $\rho$ isocontours of the non viscous 3D SPH disc modelling in a microquasar. Explicit SPH results compare with SPH implicit ones.}
\end{figure}

  Figs. 10, 11 and 12 show a comparison of explicit and implicit SPH disc structures for the non viscous and the viscous models whose $\alpha_{SS} = 0.1$ and $0.4$, respectively. All explicit and implicit structures impressively compare with each other, as well as their injection, ejection and accretion rates, whose values are in the order of $\sim 10^{21} \ g \ s^{-1}$, $6.5 \cdot 10^{20} \ g \ s^{-1}$ and $3.5 \cdot 10^{20} \ g \ s^{-1}$, respectively for the inviscid disc models, of $\sim 7.8 \cdot 10^{20} \ g \ s^{-1}$, $3.6 \cdot 10^{20} \ g \ s^{-1}$ and $4.2 \cdot 10^{20} \ g \ s^{-1}$, respectively for the $\alpha_{SS} = 0.1$ viscid disc models, of $\sim 5.9 \cdot 10^{20} \ g \ s^{-1}$, $2.6 \cdot 10^{20} \ g \ s^{-1}$ and $3.3 \cdot 10^{20} \ g \ s^{-1}$, respectively for the $\alpha_{SS} = 0.4$ viscid disc models. Some limited differences appear in the comparison between the implicit to the explicit disc structures in Figg. 10 and 11, due to the fact that the implicit structures suffer of a relative additional intrinsic viscosity, because of the cumulation of numerical dissipations in the implicit integration iterative procedure. These differences are reduced and become negligible in the physically more viscous disc structure of Fig. 12, where both the explicit and the implicit disc structures better compare with each other. In this case the higher physical viscosity makes secondary the effect of other artificial and intrinsic numerical dissipations.

\begin{figure}
\resizebox{\hsize}{!}{\includegraphics[clip=true]{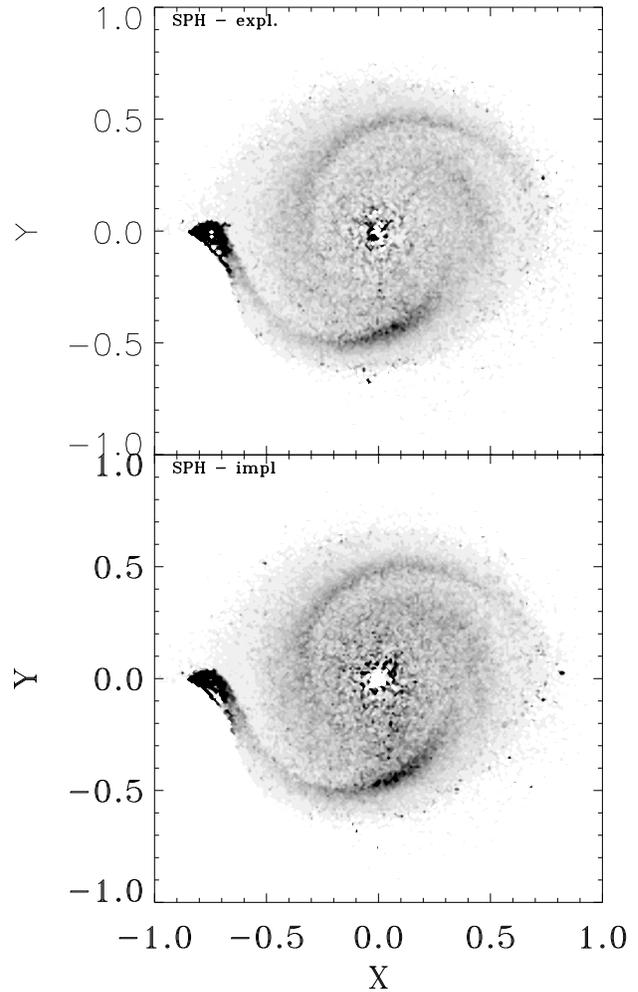}}
\caption{XY plots of $64$ greytones density $\rho$ isocontours of the $\alpha_{SS} = 0.1$ viscous 3D disc modelling in a microquasar. Explicit SPH results compare with SPH implicit ones.}
\end{figure}

  The total number of disc particles are of the order of $137613$, $143013$ and $125203$ particles, respectively in steady state conditions when the mass of the disc is statistically unchanged for explicit calculations after $522100$, $5246800$ and $7153700$ time steps for $\alpha_{SS} = 0$, $0.1$ and $0.4$, respectively. Instead, for implicit calculations, we count $135112$, $139717$ and $123150$ particles, respectively after $148300$, $178250$ and $248150$ time steps, respectively for the same $\alpha_{SS}$. As it is evident, the total number of particles compare with each other for the same $\alpha_{SS}$ characterizing the disc model, but the total number of time steps for implicit integrations is clearly much less than that relative to explicit integrations.

\begin{figure}
\resizebox{\hsize}{!}{\includegraphics[clip=true]{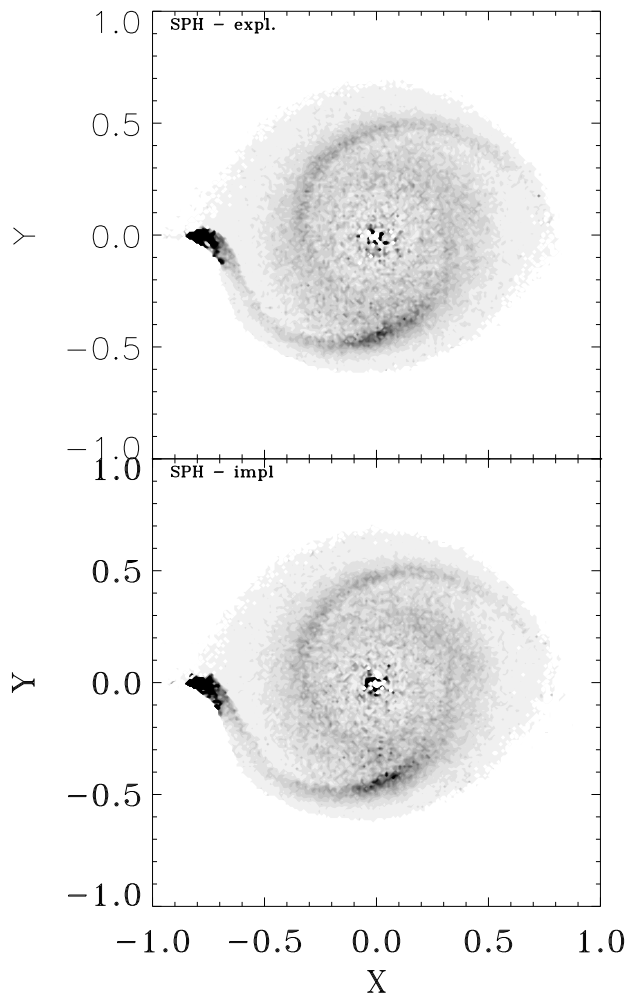}}
\caption{XY plots of $64$ greytones density $\rho$ isocontours of the $\alpha_{SS} = 0.4$ viscous 3D disc modelling in a microquasar. Explicit SPH results compare with SPH implicit ones.}
\end{figure}

  The durations of each integration process for the implicit scheme is of course longer taking into account of the iterative cycles. The total number of disc particles for implicit integration schemes are about $1 - 2 \%$ systematically lower than those relative to explicit integration schemes. Of course these are very little difference among explicit-implicit results. This result is explained by the fact that, being the mass transfer process from L1 a discrete process simulated by a particle generation, the longer implicit time step involves a mass transfer rate about $1 - 2\%$ slower in the implicit cases. Consequently, also the accretion and the ejection rates for implicit integration disc modelling are proportionally lower than those relative to the explicit integration disc models. Being these differences so tiny, they are not detectable by the direct counting of particles on short or on medium size time intervals, refereed to the orbital period. However, this very little difference in the injection, ejection, accretion rates, does not involve consistent differences among the explicit - implicit disc structures and dynamics, whose azimuthal and radial profile of density look like very similar, a part some graphical contrast effects in the grey tones.

  Even though disc models refer to a low compressibility regime ($\gamma = 1.3$), the primary's MBH Roche lobe is large and deep enough to favour a disc consistency even in the non viscous disc model. This is a first result, different from those relative to low mass binaries, where SPH models yield scarcely populated low compressibility non viscous structures \citep{c48,c57}. Although finalized to a different strategy, a statistically significant 2D structure of the disc around a MBH were obtained by \citep{c44}.

  The collisional push exerted by the flow coming from L1 on the outer edge of the disc yields an effective perturbation generating a global disc's elliptic geometry in the viscous cases. As a consequence, spiral density patterns characterize SPH viscous disc models around MBH, as well depicted in Figg. 11 and 12. Instead, possible spiral patterns are not well resolved, or they do not well develop in the SPH non viscous model because of the inadequacy of dissipation able to numerically resolve shock fronts in the low compressibility flow and because the stream flow from L1 is too sparse in order to exert an effective localized push at the disc's outer edge. In these two last figures in particular, relative to the two physically viscous models, two main spiral patterns in the density are evident, even if the glimmer of the appearance of a third one is also shown. These facts are recorded in 3D modelling. On the contrary, spiral structures around MBH, as well as spiral shocks in the disc's radial flow, are normally seen in the 2D non viscous low compressibility modelling \citep{c44}. In the case of low mass close binaries, SPH low compressibility viscous disc models are not able to show any spiral pattern because of several shortcomings in the method, especially when free edge conditions are considered \citep{c58}.

\begin{figure}
\resizebox{\hsize}{!}{\includegraphics[clip=true]{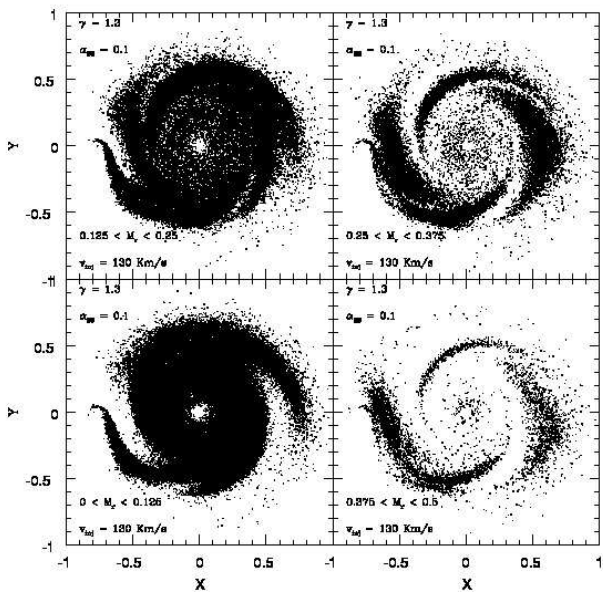}}
\caption{XY plots of the subsonic spiral kinematics ot the $\alpha_{SS} = 0.1$ viscous 3D disc in a microquasar. Selection in the radial Mach number $M_{r}$ are shown in each panel.}
\end{figure}

\begin{figure}
\resizebox{\hsize}{!}{\includegraphics[clip=true]{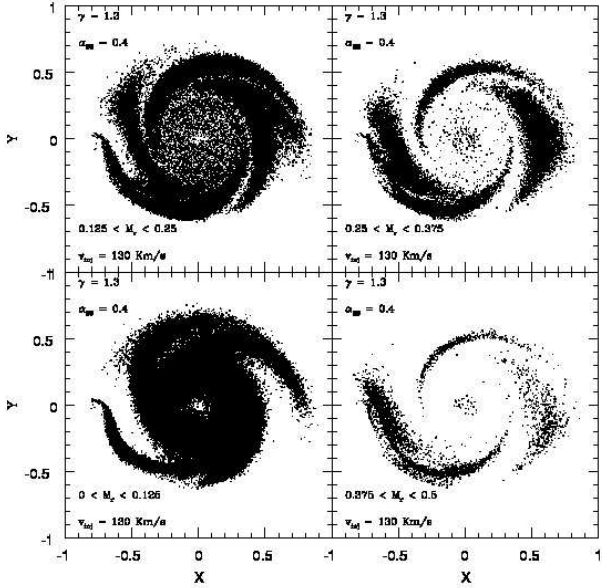}}
\caption{XY plots of the subsonic spiral kinematics ot the $\alpha_{SS} = 0.4$ viscous 3D disc in a microquasar. Selection in the radial Mach number $M_{r}$ are shown in each panel.}
\end{figure}

  The better compactness of the entire elliptical viscous disc structure, not only yields disc's spiral density profiles, but contextually also a spiral radial flow kinematics. In these low compressibility viscous models, this is a subsonic kinematics, as shown in Figg. 13 and 14 where a selection in the radial Mach number is made. The tendency should be to produce spiral shocks which should appear either increasing the flow compressibility modelling (decreasing $\gamma$) \citep{c59}, or altering the stellar mass ratio or the initial injection kinematics in order to simulate a mass transfer flow from L1 where the radial kinematics is more enhanced than the tangential one. The higher are the kinetic energy and the initial angular momentum, the better is the coming out of spiral structures and shocks \citep{c60,c61}. This is a consolidated result that has emerged for low mass binary systems. However, in the case of a microquasar, where the primary component is a MBH, a large initial angular momentum and large Coriolis and centrifugal terms favour too much an initial tangential kinematics at the cost of the radial one.

  A further comprehension of disc structure and kinematics is shown in Figg. 15 and 16, where the radial distribution of the specific angular momentum and the radial distribution of temperature are respectively shown on a logarithmic scale for the explicit integration results only because those relative to the implicit integration process are substantially the same. Two details come out from these pictures. The first one is that the inner disc regions are less populated of particles in the two viscous models. This effect, due to the enhanced radial viscous transport, makes the radial profile of these two distributions more "flat" compared with that relative to the non viscous distributions. The second particular that appear from these two pictures is that the slopes of the two radial distributions are lesser than $r^{2}\Omega \propto r^{1/2}$ and $T \propto r^{-3/4}$ relative to the "standard disc model". The temperature radial profile is even flat for a large portion of the external part of the discs. This result is a consequence of the wide geometric opening of the injected flow coming from L1 (Figg. 8 - 9), where colder and higher radial flow in the disc bulk mix with hotter flows transported from the disc outer edge toward the central accretor. As a consequence, looking from the outer disc edge toward the centre of the disc, both the temperature radial increase and the specific angular momentum decrease are lower than $T \propto r^{-3/4}$ and $r^{2}\Omega \propto r^{1/2}$ laws, even though the viscous heating effect in the disc bulk and in its inner regions is remarkable especially for the $\alpha_{SS} = 0.4$.

\begin{figure}
\resizebox{\hsize}{!}{\includegraphics[clip=true]{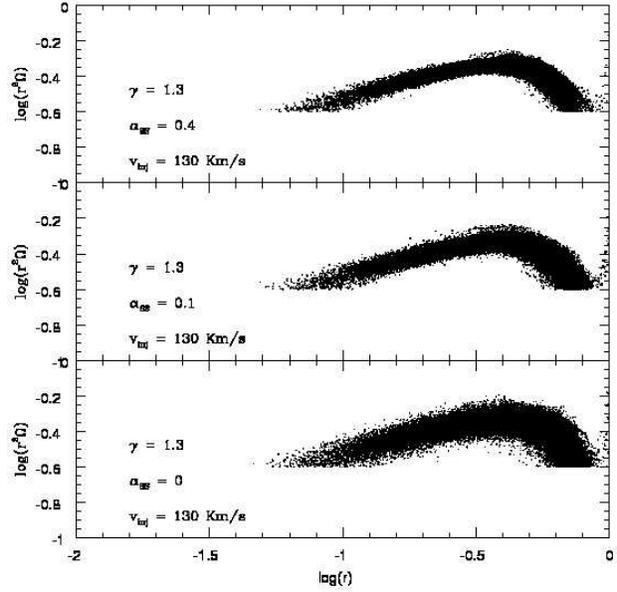}}
\caption{Radial distribution of the specific angular momentum in a logarithmic scale.}
\end{figure}

\begin{figure}
\resizebox{\hsize}{!}{\includegraphics[clip=true]{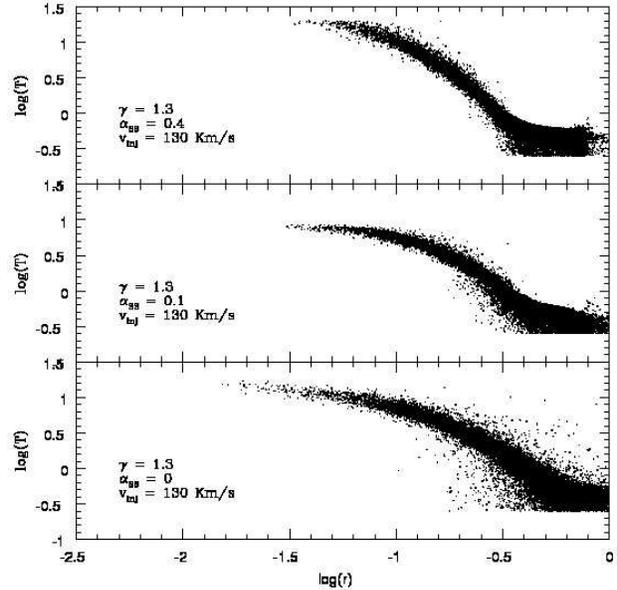}}
\caption{Radial distribution of temperature in a logarithmic scale.}
\end{figure}

  From the pure numerical point of view, the so tight correspondence among explicit and implicit disc structures means that the radial viscous transport mechanisms are comparable despite the higher accumulation of dissipation and error propagation during the implicit integration iterative cycles. This is explained by the fact that the integrated effect of dissipation, distributed on a $\Delta t_{l}$ up to $10 - 20$ times longer than $\Delta t_{SPH}$ accumulated on $3 - 4$ cycles corresponds to that relative to explicit integrations distributed on the same time interval ($\Delta t_{l} = m \Delta t_{SPH}$, where $m = \Delta t_{l}/\Delta t_{SPH}$). This non trivial numerical conclusion confirms that, at least for SPH, the chosen criterion for the number of iterative cycles works well.

\section{Concluding remarks}

  From the numerical point of view, a successful Semi-Lagrangian explicit-implicit integration numerical scheme is here applied for the SPH method that is a Free Lagrangian scheme \citep{c54}. Comparison to results obtained working with explicit numerical schemes shows an impressive convergence of results. Traditionally, numerical schemes in such explicit-implicit approaches are stable and show a convergence when the adopted implicit time step is larger than the explicit one by up to $6$ times \citep{c21,c22}. In this study, correlating our implicit time $\Delta t_{l}$ stepping to the explicit time stepping $\Delta t_{SPH}$ and to the kinematic time stepping $\Delta t_{k}$, we report consistent and stable results both for some 1D and 2D critical tests involving shocks and shockless physically radial viscous transport, as well as for 3D low compressibility accretion disc models around a MBH in a microquasar. This choice, in particular for the viscous modelling is motivated to stress the validity of the implicit technique we described. Consistency of results is recorded up to $\Delta t_{l}/\Delta t_{SPH} = 15$ as shown in Fig. 17 for the $\alpha_{SS} = 0.4$ physically viscous accretion disc modelling, where $\Delta t_{l}/\Delta t_{SPH}$ is reported as a function of time for the implicit results for the three simulated discs. This has the evident advantage that working in such an explicit-implicit scheme we can obtain meaningful results in a much shorter time than working adopting the SPH explicit time stepping. In particular we yield implicit results for the $\alpha_{SS} = 0.1$ and $\alpha_{SS} = 0.4$ discs in two weeks of cpu time using a serial code on a system based on an AMD Opteron cpu instead of times of the order of 5 months on the same hardware working in explicit integration technique. Compared with other implicit techniques, this technique, working with a larger time computational step than the $\Delta t_{SPH}$, is competitive in so far as the integrated numerical value is obtained in a few iterations, otherwise the time involved in a higher number of iterations makes the techniques not attractive. The constraint for our Semi-Lagrangian approach we propose is that the number of implicit iterations should be high enough to have stable solutions, but low enough to have accuracy with a moderate dissipation and not too time consuming. However, a fast convergence of the integrated numerical solution belongs to the SPH philosophy because of the spatial smooth distribution of any physical property around each moving particle, where $\nabla \int_D A(\bmath{r}') W(\bmath{r}, \bmath{r}', h) d \bmath{r}' \sim \int_D A(\bmath{r}') \nabla W(\bmath{r}, \bmath{r}', h) d \bmath{r}'$. At the same time, the necessity to compute some further spatial gradients according to eqs. (45, 46) does not involve any further cpu time because these spatial gradients could be calculated contextually with other SPH quantities during the explicit step described in \S 3.2.

\begin{figure}
\resizebox{\hsize}{!}{\includegraphics[clip=true]{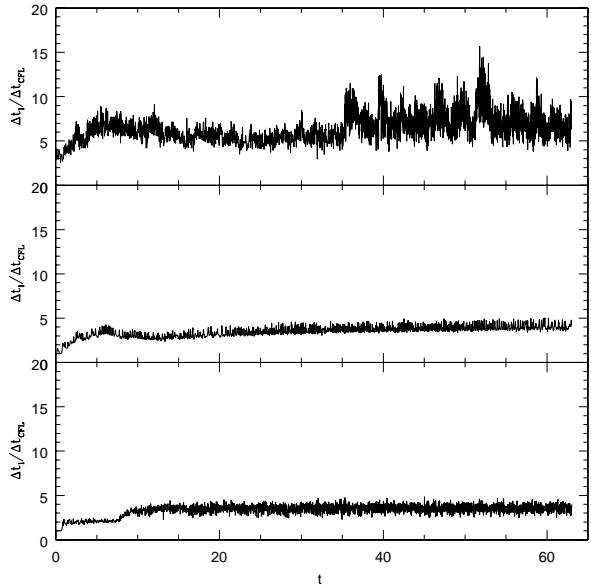}}
\caption{$\Delta t_{l}/\Delta t_{SPH}$ as a function of time $t$ for the three simulated 3D accretion discs. The panel at the bottom refers to the $\alpha_{SS} = 0$ non viscous model. $\alpha_{SS} = 0.1$ (middle panel) as well as $\alpha_{SS} = 0.4$ (top panel) results are also reported.}
\end{figure}

  On \S 3.2 we put an unanswered question on which implicit relaxed solution is obtained after numerous iterations especially when $10 < \Delta t_{l}/\Delta t_{SPH} < 20$. This argument of pure numerical applied mathematics is worth to be the theme of a dedicated book on the stability and convergence of Semi-Lagrangian implicit numerical solutions of hyperbolic - parabolic systems of equations in fluid dynamics and is well beyond the scope of this paper, whose message is that for SPH it is possible to calculate consistent Semi-Lagrangian implicit solutions even in the range $10 < \Delta t_{l}/\Delta t_{SPH} < 20$ working with $3 - 4$ iterative cycles (the first one explicit). Hence, these results further shift the limit of $6$ claimed by \citet{c21,c22} and still debated \citet{c23,c20}.

  From the astrophysical point of view, a well-bound accretion disc appears in a microquasar even for a non viscous - low compressibility modelling. This is allowed because of the strong gravitational field of the central MBH prevailing on the repulsive pressure forces. In these low compressibility models the kinematic of the radial flow stays subsonic throughout even if the physical viscous contribution is also taken into account. For the physically viscous models, two main spiral patterns in the density flow clearly appear mainly as a consequence of the elliptical morphology of the entire disc structure and the disc's outer edge is better shaped. A possible third spiral profile can also appear. The radial disc structures all deviate from that of the disc standard model because of the vertical inflow of cold gas coming from L1.

  The full consistency of the Semi-Lagrangian explicit-implicit integration technique applied to free Lagrangian methods is checked not only because of the technical tests here presented, but above all, because of the fully convergence of results between 3D SPH explicit and explicit-implicit accretion disc modelling, where a complex of phenomenology, both viscous and inviscid, are here shown, especially when a strong gravitational field of a MBH deeply affect the local viscous thermodynamics producing a consistent difference between $\Delta t_{l}$ and $\Delta t_{SPH}$. This beyond the correspondence between our disc modelling and a real existing astrophysical counterpart. Around a MBH, some physics of viscous disc models is stressed so much to determine, especially at its inner edge those conditions decreasing the explicit $\Delta t_{SPH}$ to numerical values so small to stress out if a $\Delta t_{l}/\Delta t_{SPH} > 6$ could yield to SPH consistent solutions in a Semi-Lagrangian explicit-implicit integration.

  A gravitational field, even if dominant against pressure forces is necessary to check if for large $\Delta t_{k}/\Delta t_{SPH}$ the method correctly works. Moreover, in a 3D problem, relevant gravitational forces could be dominant against pressure, viscosity ones, etc., only along the radial direction. The opposite happens along the tangential and vertical directions. This as for the momentum equation only. In any case, the consistency of solutions should also take into account of the continuity and of the thermal energy equations not including any external field. Since the complete solution must also take into account of these contributions, we deduce that the comparison of explicit and explicit-implicit solutions is a success of the new scheme.

\section*{Acknowledgments}

We thank Dr. P. Leto of the INAF - Osservatorio Astrofisico di Catania for some helpful interventions that improved the presentation of the paper.


\begin{thebibliography}{99}

\bibitem[\protect\citeauthoryear{Abolmasov \& Shakura}{2009}]{c53} Abolmasov, P., Shakura, N.I., 2009, AN, 7, 737
\bibitem[\protect\citeauthoryear{Ardeljian et al.}{1996}]{c6} Ardeljian, N.V., Bisnovatyi-Kogan, G.S., Kosmachevskii, K.V., Moiseenko, S.G., 1996, A\&ASS, 115, 573
\bibitem[\protect\citeauthoryear{Bartello and Thomas}{1996}]{b16} Bartello, P., Thomas, S.J., 1996, Mon. Wea. Rev., 124, 2883
\bibitem[\protect\citeauthoryear{Batchelor}{2000}]{c36} Batchelor, K., 2000, "An introduction to fluid dynamics", Cambridge Univ. Press
\bibitem[\protect\citeauthoryear{Boris \& Book}{1973}]{c18} Boris, J.P., Book, D.L., 1973, JCoPh, 11, 38
\bibitem[\protect\citeauthoryear{Chakrabarty}{1992}]{c59} Chakrabarty, S.K., 1992, MNRAS, 259, 410
\bibitem[\protect\citeauthoryear{Courant et al.}{1928}]{c1} Courant, R., Friedrichs, K., Lewy, H., 1928, Mathematiske Annalen, 10, 32
\bibitem[\protect\citeauthoryear{Courant et al.}{1967}]{c2} Courant, R., Friedrichs, K., Lewy, H., 1967, (Engl. transl.) IBM Journal \& AEC Report NYO-7689, 215
\bibitem[\protect\citeauthoryear{Falcone and Ferretti}{1998}]{b12} Falcone, M., Ferretti, R., 1998, SIAM J. {\it Numer. Anal.}, 35, 909
\bibitem[\protect\citeauthoryear{Flebbe et al.}{1994a}]{c14} Flebbe, O., M\"unzel, H., Riffert, H., Herold, H., 1994a, Mem. S.A.It, 65, 1049
\bibitem[\protect\citeauthoryear{Flebbe et al.}{1994b}]{c15} Flebbe, O., M\"unzel, H., Herold, H., Riffert, H., Ruder, H., 1994b, ApJ, 431, 754
\bibitem[\protect\citeauthoryear{Fletcher}{1991}]{c3} Fletcher, C.A.J., 1991, "Computational techniques for fluid dynamics", Springer
\bibitem[\protect\citeauthoryear{Gibbs}{1898}]{c40} Gibbs, J.W., 1898, Nature, 59, 200
\bibitem[\protect\citeauthoryear{Gibbs}{1899}]{c41} Gibbs, J.W., 1898, Nature, 59, 606
\bibitem[\protect\citeauthoryear{Gravouil \& Comberscure}{2001}]{c31} Gravouil, A., Comberscure, A., 2001, IJNME, 50, 199
\bibitem[\protect\citeauthoryear{Hirsch}{1997}]{c4} Hirsch, C., 1997, "Numerical computation of internal and external flows", Wiley
\bibitem[\protect\citeauthoryear{Ismail et al.}{2009}]{c29} Ismail, F., Ken, Y.L., Othman, M., 2009, IJMA, 3, 239
\bibitem[\protect\citeauthoryear{Ketcheson et al.}{2009}]{c26} Ketcheson, D.I., Macdonald, C.B., Gottlieb, S., 2009, JANM, 59, 373
\bibitem[\protect\citeauthoryear{King et al.}{2007}]{c52} King, A.R., Pringle, J.E., Livio, M., 2007, MNRAS, 376, 1740
\bibitem[\protect\citeauthoryear{Lanzafame}{2008}]{c45} Lanzafame, G. 2008, PASJ, 60, 259
\bibitem[\protect\citeauthoryear{Lanzafame}{2009}]{c46} Lanzafame, G., 2009, AN, 330, 843
\bibitem[\protect\citeauthoryear{Lanzafame}{2010a}]{c16} Lanzafame, G., 2010a, MNRAS, 408, 2336
\bibitem[\protect\citeauthoryear{Lanzafame}{2010b}]{c58} Lanzafame, G., 2010b, MNRAS, 408, 1551
\bibitem[\protect\citeauthoryear{Lanzafame}{2010c}]{c24} Lanzafame, G., 2010c, ASP, 429, 106
\bibitem[\protect\citeauthoryear{Lanzafame \& Belvedere}{2001}]{c42} Lanzafame, G., Belvedere, G., 2001, JKAS, 34, S313
\bibitem[\protect\citeauthoryear{Lanzafame \& Belvedere}{2002}]{c43} Lanzafame, G., Belvedere, G., 2002, PASJ, 54, 781
\bibitem[\protect\citeauthoryear{Lanzafame \& Belvedere}{2005}]{c44} Lanzafame, G., Belvedere, G., 2005, ApJ, 632, 499
\bibitem[\protect\citeauthoryear{Lanzafame et al.}{1992}]{c57} Lanzafame, G., Belvedere G., Molteni D., 1992, MNRAS, 258, 152
\bibitem[\protect\citeauthoryear{Lanzafame et al.}{2006}]{c47} Lanzafame, G., Belvedere G., Molteni, D., 2006, A\&A, 453, 1027
\bibitem[\protect\citeauthoryear{Lanzafame et al.}{2011}]{c25} Lanzafame, G. Costa, V., Belvedere, G., 2011, ASP, in the press
\bibitem[\protect\citeauthoryear{Lanzafame et al.}{2000}]{c60} Lanzafame, G., Maravigna, F., Belvedere G., 2000, PASJ, 52, 515
\bibitem[\protect\citeauthoryear{Lanzafame et al.}{2001}]{c61} Lanzafame, G., Maravigna, F., Belvedere G., 2001, PASJ, 53, 139
\bibitem[\protect\citeauthoryear{Lattanzio et al.}{1985}]{c17} Lattanzio, J.C., Monaghan J.J., Pongracic, H., Schwarz, M.P., 1985, MNRAS, 215, 125
\bibitem[\protect\citeauthoryear{LeVeque}{1992}]{c37} LeVeque, R.J., 1992, "Numerical methods for conservation laws", Lectures in Mathematics, ETH Z\"urich, Birkh\"auser
\bibitem[\protect\citeauthoryear{LeVeque}{2002}]{c5} LeVeque, R.J., 2002, "Finite volume methods for hiperbolic problems", Cambridge Univ. Press
\bibitem[\protect\citeauthoryear{Lubow \& Shu}{1975}]{c49} Lubow, S.H., Shu, F.H.,  1975, MNRAS, 198, 383
\bibitem[\protect\citeauthoryear{Majid et al.}{2006}]{c28} Majid, Z.A., Suleiman, M.B., Omar, Z., 2006, BMMSS, 29, 23
\bibitem[\protect\citeauthoryear{Meglicki et al.}{1993}]{c19} Meglicki, Z., Wickramasinghe, D., Bicknell, G.V., 1993, MNRAS, 264, 691
\bibitem[\protect\citeauthoryear{Miranda et al.}{1989}]{c30} Miranda, I., Ferencz, R.M., Hughes, T.J.R., 1989, EESD, 18, 643
\bibitem[\protect\citeauthoryear{Molteni et al.}{1991}]{c48} Molteni, D., Belvedere, G., Lanzafame, G., 1991, MNRAS, 249, 748
\bibitem[\protect\citeauthoryear{Monaghan}{1985}]{c11} Monaghan, J.J., 1985, Comp. Phys. Rept., 3, 71
\bibitem[\protect\citeauthoryear{Monaghan}{1992}]{c12} Monaghan, J.J., 1992, ARA\&A, 30, 543
\bibitem[\protect\citeauthoryear{Monaghan}{1997}]{c39} Monaghan, J.J., 1997, JCoPh, 136, 298
\bibitem[\protect\citeauthoryear{Monaghan \& Lattanzio}{1985}]{c13} Monaghan, J.J., Lattanzio, J.C., 1985, A\&A, 149, 135
\bibitem[\protect\citeauthoryear{Mosqueda \& Ahmadizadeh}{2010}]{c35} Mosqueda, G., Ahmadizadeh, M., 2010, EESD, doi: 10. 1002/eqe. 1066
\bibitem[\protect\citeauthoryear{Petkova \& Springel}{2009}]{c7} Petkova, M., Springel, V., 2009, MNRAS, 396, 1383
\bibitem[\protect\citeauthoryear{Pirroneau}{1982}]{b13} Pirroneau, O., 1982, Num. Math, 38, 309
\bibitem[\protect\citeauthoryear{Press et al.}{1992}]{c56} Press, W.H., Teukolsky, S.A., Vettering, W.T., Flannery, B.P., 1992, "Numerical Recipes" (2nd ed.; Cambridge Univ. Press)
\bibitem[\protect\citeauthoryear{Pringle}{1981}]{b3} Pringle, J.E., 1981, ARA\&A, 19, 137
\bibitem[\protect\citeauthoryear{Robert}{1969}]{c21} Robert, A., 1969, "The integration of a spectral model of the atmosphere by the implicit method", Proc. of WMO/IUGG Symp. on NWP, Tokyo, Japan Metereological Agency, VII, 19-VII.24
\bibitem[\protect\citeauthoryear{Robert}{1981}]{c23} Robert, A. 1981, Atmos. Ocean., 19, 35
\bibitem[\protect\citeauthoryear{Robert et al.}{1972}]{c22} Robert, A., Henderson, J., Turnbull, C., 1972, MWR, 100, 329
\bibitem[\protect\citeauthoryear{Rook et al.}{2007}]{c8} Rook, R., Yildiz, M., Dost, S., 2007, Num. Heat Trans. B., 51, 1
\bibitem[\protect\citeauthoryear{Sahin \& Owens}{2003}]{c33} Sahin, M., Owens, R., 2003, IJNMF, 42, 57
\bibitem[\protect\citeauthoryear{Sahin \& Owens}{2006}]{c34} Sahin, M., Owens, R., 2003, IJNMF, 42, 79
\bibitem[\protect\citeauthoryear{Sewell}{1988}]{c70} Sewell, G., 1988, ``The Numerical Solution of Ordinary and Partial Differential Equations'', Academic Press
\bibitem[\protect\citeauthoryear{Shakura}{1972}]{c50} Shakura, N.I. 1972, Astron. Zh., 49, 921. (English tr.: 1973, Sov. Astron., 16, 756)
\bibitem[\protect\citeauthoryear{Shakura \& Sunyaev}{1973}]{c51} Shakura, N.I., Sunyaev, R.A., 1973, A\&A, 24, 337
\bibitem[\protect\citeauthoryear{Shampine}{1994}]{c71} Shampine, L.F., 1994, ``Numerical Solution of Ordinary Differential Equations'', Chapman \& Hall
\bibitem[\protect\citeauthoryear{Speith \& Kley}{2003}]{b11} Speith, R., Kley, W., 2003, A\&A, 399, 395
\bibitem[\protect\citeauthoryear{S\"uli}{1988}]{b14} S\"uli, E., 1988, Num. Math., 53, 459
\bibitem[\protect\citeauthoryear{Susa}{2006}]{c9} Susa, H., 2006, PASJ, 58, 445
\bibitem[\protect\citeauthoryear{Staniforth \& C\'ot\'e}{1991}]{c20} Staniforth, A., C\'ot\'e, J., 1991, MWR, 119, 2206
\bibitem[\protect\citeauthoryear{Toro}{1999}]{c38} Toro, E.G., 1999, "Riemann solvers and numerical methods for fluid dynamics", Springer-Verlag
\bibitem[\protect\citeauthoryear{Viau et al.}{2006}]{c55} Viau, S., Bastien, P., Cha, Seung-Hoon, 2006, ApJ, 639, 559
\bibitem[\protect\citeauthoryear{Visbal \& Gaitonde}{1999}]{c27} Visbal, M.R., Gaitonde, D.V., 1999, AIAA, 10, 1231
\bibitem[\protect\citeauthoryear{Whitehouse \& Bate}{2004}]{c10} Whitehouse, S.C., Bate, M.R., 2004, MNRAS, 353, 1078
\bibitem[\protect\citeauthoryear{Whitehurst}{1995}]{c54} Whitehurst, R., 1995, MNRAS, 277, 655
\bibitem[\protect\citeauthoryear{Xiu and Em Karniadakis}{2001}]{b15} D. Xiu, Em Karniadakis, G., 2001, JcoPh., 172, 658
\bibitem[\protect\citeauthoryear{Ying et al.}{2008}]{c32} Ying, W., Rose, D.J., Henriquez, C.S., 2008, IEEE Trans. on Biomed. Eng., 55, 2701

\end{thebibliography}
\end{document}